\documentclass[a4paper,fleqn,usenatbib]{mnras}

\usepackage[T1]{fontenc}
\usepackage{ae,aecompl}
\usepackage{soul}

\usepackage{graphicx}	
\usepackage{amsmath}	
\usepackage{amssymb}	

\renewcommand{\vec}[1]{\boldsymbol{#1}}
\renewcommand{\overrightarrow}[1]{\boldsymbol{#1}}
\renewcommand{\Re}{\mathrm{Re}}
\newcommand{\tensor}[1]{\hat{\boldsymbol{#1}}}
\newcommand{\const}{\mathrm{const}}

\title[Precession of neutron stars]{A possible way to reconcile long-period precession with vortex pinning in neutron stars}

\author[O. A. Goglichidze and D. P. Barsukov]{
O. A. Goglichidze \thanks{E-mail: goglichidze@gmail.com} and D. P. Barsukov
\\
Ioffe Institute, Saint~Petersburg 194021, Russian~Federation
}

\date{Accepted XXX. Received YYY; in original form ZZZ}

\pubyear{2019}

\begin{document}
\label{firstpage}
\pagerange{\pageref{firstpage}--\pageref{lastpage}}
\maketitle

\begin{abstract}
 We propose a possible way to solve the problem of inconsistency between the neutron star long-period precession and superfluid vortex pinning, which is the basis of the most successful theories of pulsar glitches.  We assume that the pinning takes place in the region of the neutron star core, which, being magnetically decoupled, can rotate relative to the crust.
 In the framework of a simple three-component model we show that these two phenomena can coexist in the same pulsar.
 Some constraints on the formally introduced interaction coefficients following from observation data are formulated. 
\end{abstract}




\section{Introduction}

  Radio pulsars are the sources of exceptionally stable pulse sequences with very slowly decreasing frequency. 
  Some pulsars, besides smooth slow-down, demonstrate so called glitches, which are  sudden increases of the frequency and its time derivative followed by smooth recovery towards pre-glitch values.	
  The relative amplitude of the glitches $\Delta \Omega/\Omega$ lies in the range of $ \sim 10^{-12} - 10^{-5}$ \citep{EspinozaLyneStappersKramer2011}\footnote{see also http://www.jb.man.ac.uk/pulsar/glitches.html} while the recovery time-scale is of the order of days to months \citep{LyneShemarSmith2000}.
  The long time-scales of the glitch recovery indicate that the nucleon superfluidity supposed to be present in neutron star interiors participates in this phenomenon \citep{BaymPethickPines1969b}.
  The most successful  theories of glitches are based on the assumption that the neutron superfluid vortices are pinned in some region of the neutron star interior \citep{HaskellMelatos2015}. 
  The pinned superfluid conserves angular momentum and, from time to time, releases part of it, 
  spinning up the outer component of the star.
  
  
  There are several pieces of evidence that isolated neutron stars can precess with long periods.
  Pulsar  B1821-11 demonstrates correlated periodic variations of spin-down rate and beam shape \citep{StairsLyneShemar2000}.
  The most favorable explanation for this is the precession of the neutron star with period $T_p \approx 500$ d \citep{AshtonJonesPrix2016}.
  Several pulsars show periodic variations of  spin-down rate without significant correlation with beam shape \citep{KerrHobbsJohnstonShannon2016}. The time-scales of the variations are  0.5 -- 1.5~yr. 
  Some pulsars switch between two spin down states. Despite the fact that the switching itself is a fast process ( < 1 min), there are at least six pulsars for which the probability of being in a particular state is a quasi-periodic function with a characteristic time-scale $\sim 10^2$ days \citep{LyneEtAl2010}. Hence, it is natural to assume that this probability depends on the precession phase \citep{Jones2012}.
  
  Besides the directly observed variations, there are observational data, which can also be interpreted as manifestations of neutron star precession but with much longer periods.
  \citet{LyneEtAl2013} found a steady increase in separation between the main pulse and the interpulse of Crab pulsar at 0.62$^\circ \pm 0.03^\circ$ per century.
  The authors concluded that this is a consequence of the increase of the pulsar inclination angle.
  The rate of increase seems to be too large for secular  evolution but it can be ensured by free precession with period $\sim 10^2$ yr \citep{ArzamasskiyEtAl2015}.
  \citet{BiryukovBeskinKarpov2012} argued that anomalously large braking indices indicate that the pulsar spin-down rate can oscillate at the time-scale of $10^3-10^4$ years. This oscillation can be caused by precession with corresponding periods.
   The stellar magnetic field by itself should make neutron stars precess at such time-scales \citep{Melatos2000}.
   
  The problem is that it is difficult for neutron vortex pinning and  long-period ($\gtrsim $1 year) precession to coexist in the same neutron star.  
  As it was first pointed out by \citet{Shaham1977}, 
  the pinning of superfluid vortices dramatically decreases the period of precession.
  {Thus, on the one hand, the glitch theories require about 1 percent of the total stellar moment of inertia to be contained in a pinned superfluid \citep{AnderssonGlampedakisHoEspinoza2012}.
  On the other hand, if a neutron star precess with periods $\sim$1 yr or longer,  only part of $<10^{-8}$ of the total moment of inertia can be in the pinned superfluid.}
  Shaham considered a  simple model with perfectly pinned vortices. 
  Several attempts to attack this problem with more detailed models of pinned vortex	 dynamics \citep{SedrakianWassermanCordes1999,LinkCutler2002,Alphar2005}
  or assuming that pinning is absent  \citep{Link2006,KitiashviliGusev2008}
  have been made. 
  However, to date the problem remains current \citep{JonesAshtonPrix2017}.    
  

  In the present paper we suggest a way that may allow these two phenomena to be to reconciled.
  The paper is organized as follows. In Section \ref{sec:rigid_body} we introduce basic notations and formulate the problem. 
  {In Section \ref{sec:two_comps_cs} we perform a linear  mode analysis for a two-component neutron star with the superfluid pinned in the crust.
  Section \ref{sec:two_comps_gs} contains the same analysis but for a system with the superfluid located in the core.
  }
  In Section \ref{sec:three_comps} we formulate a full  three-component model and perform the linear mode analysis for it.
  Section \ref{sec:qstat_evolution} is devoted to the derivation of quasi-stationary equations, allowing the long-term evolution of the precessing neutron stars to be studied.
  In Section \ref{sec:glitch} we study the ability of the model to show glitch-like events and what restrictions for formally introduced coefficients can be obtained from the glitch observation data.
  In Discussion we speculate on the possible physical realization of the model, discuss some observational data and draw conclusions.

\section{Formulation of the problem}
  \label{sec:rigid_body}
  Let us consider a rotating neutron star.
  From the point of view of an observer in an inertial frame of reference the following equation should be satisfied: 
  \begin{equation}
    \label{eq:dM_rigid_body}
     d_t{\vec{M}} = \vec{K},
  \end{equation}
  where $\vec{M}$ is the stellar angular momentum and $\vec{K}$ is external electromagnetic torque.
  
  We start with the the ``rigid-body'' approximation assuming that the whole star rotates with a uniform angular velocity that  we denote by $\vec{\Omega}_c$. The index ``$c$'' is introduced for the sake of notation unification with the subsequent sections of this paper. 
  We assume that the star has an axisymmetric shape maintained by the crust rigidity \citep{CutlerUshomirskyLink2003} or by the magnetic field \citep{Wasserman2003}.  
  In this case, the angular momentum can be represented in the following form:
  \begin{equation}
    \label{eq:M_rigid_star}
    \vec{M} = I \vec{\Omega}_c + I \epsilon (\vec{\Omega}_c\cdot\vec{e}_c)\vec{e}_c,
  \end{equation}
  where $I$ is the star's moment of inertia, $\vec{e}_c$ is the unit vector directed along the star's symmetry axis and $\epsilon$ is the oblateness parameter.

  To date, there is no consensus concerning the exact form of external torque $\vec{K}$ acting on isolated neutrons stars \citep{BeskinChernovGwinnTchekhovskoy2015}. However, it is generally accepted that it consists of two parts of  different nature
    {\citep{DavisGoldstein1970,GoodNg1985,Melatos2000}}: 
  \begin{equation}
    \label{eq:K} 
     \vec{K} = \vec{K}_2 + \vec{K}_3,
  \end{equation}
  where
  {$\vec{K}_2 \propto \Omega_c^2$}
  is the so-called anomalous torque originating from the inertia of the near-zone electromagnetic field 
  \citep{GoodNg1985,BeskinZheltoukhov2014,GBT2015}
  while 
  {$\vec{K}_3 \propto \Omega_c^3$} is related to the angular momentum transfer away from the star to infinity by both particles and electromagnetic radiation \citep{DavisGoldstein1970,Jones1976,BeskinGurevichIstomin1983,BarsukovPolyakovaTsygan2009,PhilippovTchekhovskoyLi2014}.
  
  {Relation  ${K}_3/{K}_2$ can be estimated as
  $\sim (r_\mathrm{ns}\Omega_c/c)$,
  i.e.
  it is small  for the most pulsars (except the fastest millisecond ones, which are not considered in the present paper)
  We will ignore term $\vec{K}_3$ in Sections \ref{sec:rigid_body}-\ref{sec:three_comps} restricting ourselves to the consideration of time-scales much smaller than $\tau_x \sim \Omega_c/K_3$. The effects of this term are discussed  in  Section \ref{sec:qstat_evolution}.

   
  In the case of an axisymmetric stellar magnetic field,  anomalous torque $\vec{K}_2$  can be represented in the following form \citep{GBT2015}:
  \begin{equation}
    \vec{K}_2 = -\delta I_m (\vec{\Omega}\cdot\vec{e}_{m})[\vec{\Omega}\times\vec{e}_m],
  \end{equation}
  where $\vec{e}_m$ is the unit vector directed along the field symmetry axis,
  \begin{equation}
    \delta I_m \sim 
    10^{32} r_6^5 B_{12}^2 \mbox{ g cm}^2,
  \end{equation}
  $r_6$ is the neutron star radius in units of $10^6$ cm and $B_{12}$ is the surface magnetic field in units of $10^{12}$G.
  In the special situation of vector $\vec{e}_m$ coinciding with vector $\vec{e}_c$, the anomalous torque can be taken into account just by redefining the oblateness parameter \citep{ZanazziLai2015,GBT2015}: 
  \begin{equation}
    \label{eq:epsilon_new}
    \epsilon_\mathrm{new} =\epsilon_\mathrm{old} + \delta I_m/I,
  \end{equation} 
  where $\epsilon_\mathrm{old}$ characterizes the stellar matter deformarion only.
  In general case,
  the anomalous torque can be taken into account by modification of star moment of inertia tensor. 
  
  
  Ignoring $\vec{K}_3$ and including $\vec{K}_2$ in $\vec{M}$ (without changing the notation), we can reduce equation \eqref{eq:dM_rigid_body} to 
  \begin{equation}
    \label{eq:dM_rigid_body_free}
     d_t \vec{M}  = 0.
  \end{equation}
  In Sections \ref{sec:rigid_body}-\ref{sec:three_comps} we will consider only the axisymmetric case. Therefore, formula \eqref{eq:M_rigid_star} is assumed to be valid, where the oblateness parameter is now calculated with formula \eqref{eq:epsilon_new}.  The more general case of a triaxial star is discussed in Section \ref{sec:qstat_evolution}.
  
  Equation \eqref{eq:dM_rigid_body_free} formally states the conservation of total angular momentum. 
  It is easy to verify that the mechanical energy $E = (\vec{\Omega}\cdot\vec{M})/2$ is also conserved.
  It is natural because we neglect torque $\vec{K}_3$ describing the energy and angular momentum flows away from the star.
  The effects of this term are discussed  in  Section \ref{sec:qstat_evolution}.
  
}  
  Substituting angular momentum expression \eqref{eq:M_rigid_star} into equation \eqref{eq:dM_rigid_body_free} and changing to the co-rotating frame of reference, we obtain
  \begin{equation}
    \label{eq:dOm_rigid_body}
    \dot{\vec{\Omega}}_c = \epsilon  (\vec{e}_c\cdot\vec{\Omega}_c) [\vec{e}_c \times \vec{\Omega}_c].
  \end{equation}
  Here and further we will denote by a superscript dot the time derivative in the frame of reference rotating with angular velocity $\vec{\Omega}_c$.
  The solution to this equation is the uniform rotation of vector $\vec{\Omega}_c$ about symmetry axis $\vec{e}_c$ with angular frequency
  \begin{equation}
    \label{eq:omega_p_rigid_body}
    \omega_p = \epsilon\cos\theta\Omega_c,
  \end{equation}
  { where $\theta$ is the angle between angular velocity $\vec{\Omega}_c$ and symmetry axis $\vec{e}_c$.} 
  This type of motion is called free precession. 
  {
  Introducing pulsar rotational period $P = {2\pi}/{\Omega_c}$ and period of pulsar precession $T_p={2\pi}/{\omega_p}$, directly from \eqref{eq:omega_p_rigid_body} we have
  \begin{equation}
    \label{eq:Tp_rigid_body}
    T_p = \frac{P}{\epsilon \cos\theta}.
  \end{equation}
  }
  
  Equation \eqref{eq:dOm_rigid_body} has a fixed point corresponding to stationary rotation about the symmetry axis:
  \begin{equation}
    \vec{\Omega}_c = \vec{\Omega}_0 =\const, \ \ \ \vec{\Omega}_0\parallel \vec{e}_c.
  \end{equation}
  In order to study perturbations to the stationary rotation let us introduce a departure vector: 
  \begin{equation}
    \label{eq:mu_rigid_star}
    \vec{\mu}_c = \vec{\Omega}_c - \vec{\Omega}_0.
  \end{equation}
  Further, it will be convenient to use the following notations, which can be applied to an arbitrary vector $\vec{V}$:
  \begin{equation}
    V^\parallel = (\vec{V}\cdot\vec{e}_z), \ \ \ V^\perp  = (\vec{V}\cdot\vec{e}_x) + i (\vec{V}\cdot\vec{e}_y),
  \end{equation}
  where $(\vec{e}_x,\vec{e}_y,\vec{e}_z)$ is an orthonormal basis fixed in co-rotating frame of reference such that $\vec{e}_z = \vec{e}_c$.
  Substituting \eqref{eq:mu_rigid_star} into \eqref{eq:dOm_rigid_body} and neglecting the quadratic in ${\mu}_c^\parallel$ and ${\mu}_c^\perp$ terms, we obtain
  \begin{align}
    \label{dmu_par_rigid_star}
    &\dot{\mu}_c^\parallel = 0, \\
    \label{dmu_perp_rigid_star}
    &\dot{\mu}_c^\perp = i \epsilon \Omega_0 \mu_c^\perp.
  \end{align}
  
  Free precession motion is described by the ``perpendicular'' equation \eqref{dmu_perp_rigid_star}. Substituting $\mu_c^\perp \propto \exp(p t)$
  we find
  \begin{equation}
    p = i\epsilon \Omega_0.
  \end{equation}
  Hence, 
  the perpendicular part of vector $\vec{\mu}$ rotates about the symmetry axis with angular frequency $\omega_p = \epsilon \Omega_0$, which in the linear approximation ($\cos\theta \approx 1$, $\Omega_c \approx \Omega_0$) coincides with expression \eqref{eq:omega_p_rigid_body}.
  Thus, knowing pulsar period $P$ and period of pulsar precession $T_p$, one can estimate the oblateness parameter as
  \begin{equation}
    \epsilon \sim \frac{P}{T_p}.
  \end{equation}

  As it was first pointed out by \citet{Shaham1977}, the situation dramatically changes if a neutron star contains a pinned superfluid.
  In this case, the stellar angular momentum expression \eqref{eq:M_rigid_star} requires some modification, namely, 
  \begin{equation}
    \label{eq:M_rigid_star_wpin}
    \vec{M} = \tilde{I} \vec{\Omega}_c + \epsilon \tilde{I}  (\vec{\Omega}_c\cdot\vec{e}_c)\vec{e}_c + \vec{L}_\mathrm{sf}.
  \end{equation}
  where $\vec{L}_\mathrm{sf}$ is the angular momentum of the pinned superfluid and $\tilde{I}$ is the moment of inertia of the rest of the stellar matter (excluding the pinned superfluid).
  
  The angular momentum of the superfluid is totally determined by the distribution of quantum vortices. If the vortices are pinned, their density and orientation are fixed in the frame of reference rotating with angular velocity $\vec{\Omega}_c$.
  Therefore, angular momentum $\vec{L}_\mathrm{sf}$ is constant in the co-rotating frame of reference:
  \begin{equation}
    \label{eq:dL_rigid_body}
    \dot{\vec{L}}_\mathrm{sf} = 0.
  \end{equation}
 
  For the sake of simplicity we restrict ourselves to the consideration  of a particular configuration of $\vec{L}_\mathrm{sf}\parallel\vec{e}_c$. Such a configuration is not unrealistic but it is far from the general case.
  The analysis of star rotation with arbitrarily oriented $\vec{L}_\mathrm{sf}$ was given by \citet{Shaham1977}. It shows no significantly different results.
  
  Substituting expressions \eqref{eq:M_rigid_star_wpin} and \eqref{eq:dL_rigid_body} into equation \eqref{eq:dM_rigid_body_free} and changing to the co-rotating frame of reference, we obtain
  \begin{equation}
    \label{eq:dOm_rigid_body_wpin}
    \dot{\vec{\Omega}}_c = \left(\epsilon + \frac{L_\mathrm{sf}}{\tilde{I}\Omega_c\cos\theta}\right) \Omega_c \cos\theta  [\vec{e}_c \times \vec{\Omega}_c].
  \end{equation}
  At first sight this equation
   is similar to
  equation \eqref{eq:dOm_rigid_body}.
  However, the rate of vector $\vec{\Omega}_c$ rotation is now determined by the sum of the two terms in parentheses.
  The second term can be estimated as $\sim I_\mathrm{pin}/\tilde{I}$, where $I_\mathrm{pin}$ is the moment of inertia of the pinned superfluid. 
  If the understanding of glitches as a manifestation of neutron vortex dynamics 
  is correct,   at least about 1 percent of star's moment of inertia should be contained in the pinned superfluid ($I_\mathrm{pin}/\tilde{I} \gtrsim 10^{-2}$) to ensure the observed glitch activity of the pulsars \citep{AnderssonGlampedakisHoEspinoza2012}.
  { As for the first term, $\epsilon$, it hardly can reach such large magnitudes. Calculations give values at least several orders of magnitude smaller \citep{CutlerUshomirskyLink2003,Wasserman2003}. Hence, the second term in parentheses is dominant.}  

  After  linearization we obtain
  \begin{align}
    \label{dmu_par_rigid_star_wp}
    &\dot{\mu}_c^\parallel = 0, \\
    \label{dmu_perp_rigid_star_wp}
    &\dot{\mu}_c^\perp = i \tilde{\epsilon} \Omega_0 \mu_c^\perp,
  \end{align}  
  where we have introduced the effective oblateness parameter 
  \begin{equation}
    \tilde{\epsilon} = \epsilon + \frac{L_\mathrm{sf}}{\tilde{I}\Omega_0} \approx I_\mathrm{pin}/\tilde{I} \gtrsim 10^{-2}.
  \end{equation}  
  Substituting  $\mu_c^\perp \propto \exp(p  t)$ into equation \eqref{dmu_perp_rigid_star_wp}, we find
  \begin{equation}
    \label{eq:prec_mode_rigid_star_wp}
    p = i\tilde{\epsilon} \Omega_0 
  \end{equation}   
  
    {Thus, one can see that the free precession is governed almost exclusively by the amount of pinned superfluid which does not allow the star to precess with periods substantially longer than }
  \begin{equation}
     T_p \lesssim  \frac{P}{\tilde{\epsilon}} \sim 10^2 P. 
  \end{equation}
  Such short periods are many orders of magnitude smaller than the observed values.
  
    
  

\section{Two-component system}
  \label{sec:two_comps_cs}

   In the previous section we assumed that the whole star except the pinned superfluid rotates as a rigid body.
  Let us relax this assumption.
  First we consider the simplest case supposing that the neutron star consists of two dynamically distinguished components: an outer c-component whose rotation is directly observed and an inner g-component \citep{Shaham1977,SedrakianWassermanCordes1999}. At this point, the components are introduced phenomenologically. The indices  ``c'' and ``g''  are chosen for the sake of unification with the full model presented in Section \ref{sec:three_comps}. 
 {For the components we can write}
  \begin{align}
    \label{eq:dMc_two_comp}
    &d_t {\vec{M}}_c = \vec{N}_{gc}, \\
    \label{eq:dMg_two_comp}
    &d_t {\vec{M}}_g = \vec{N}_{cg},
  \end{align}
  where 
  {
  $\vec{M}_i$ are the components angular momenta and $\vec{N}_{ij}$ are the components interaction torques, $i,j = c, g$.
  The c-component consists of the ``normal" fraction and the pinned superfluid fraction.
  By the ``normal" fraction we mean all the possible constituents of the c-component except the pinned neutron superfluid. 
  It includes the ions in the crust lattice sites, electrons and possible non-superfluid neutrons. It could  also include the part of stellar core matter that is rigidly connected to the crust.
  The ``normal" fraction is assumed  to be axisymmetric with moment of inertia $I_c$ and oblateness parameter $\epsilon_c$. It rotates with angular velocity $\vec{\Omega}_c$.
  The pinned superfluid contains angular momentum $\vec{L}_\mathrm{sf}$ directed along the ``normal" fraction symmetry axis $\vec{e}_c$.
  Thus, the total c-component angular momentum has the form
  \begin{equation}
    \label{eq:Mc_two_comp}
    \vec{M}_c = {I}_c \vec{\Omega}_c + {I}_c \epsilon_c (\vec{\Omega}_c\cdot\vec{e}_c)\vec{e}_c + \vec{L}_\mathrm{sf}.
  \end{equation}
  
  The g-component is the part of the core that is not rigidly connected to the crust. This component, for simplicity, is assumed to be spherically symmetric.
  It is characterized by moment of inertia $I_g$ and rotates with angular velocity $\vec{\Omega}_g$. Therefore, its angular momentum is equal to 
  \begin{equation}
    \label{eq:Mg_two_comp}
    \vec{M}_g = {I}_g \vec{\Omega}_g.
  \end{equation}
  
  Since the total stellar angular momentum $\vec{M} = \vec{M}_c+\vec{M}_g$ should be conserved,  $\vec{N}_{cg} = - \vec{N}_{gc}$.
  However, for the mechanical energy we have
  \begin{equation}
    d_t E = -(\vec{\Omega}_c-\vec{\Omega}_g)\cdot\vec{N}_{cg}.
  \end{equation}
  Therefore, the non-zero components of the interaction torque parallel to angular velocity difference $(\vec{\Omega}_c-\vec{\Omega}_g)$ lead to the dissipation of mechanical energy inside the star.
  }
  
  Substituting expressions \eqref{eq:Mc_two_comp} and \eqref{eq:Mg_two_comp} into equations \eqref{eq:dMc_two_comp} and \eqref{eq:dMg_two_comp},
  changing to the frame of reference co-rotating with the c-component, and taking into account equation \eqref{eq:dL_rigid_body}, we obtain the equations for angular velocities:
  \begin{align}
     \label{dOmega_c_two_comps_cs}
     &\dot{\vec{\Omega}}_c  + \left(\epsilon_c + \frac{\omega_\mathrm{sf}}{\Omega\cos\theta}\right)(\vec\Omega_c\cdot \vec{e}_c)  [\vec{\Omega} \times \vec{e}_c ]  \\
     \nonumber
     & \ \ \ \ \ \ \ \ \ \ \ \ \ \  \ \ \ \ \ \ \ \ \ \ \ \ \ =\vec{R}_{gc}  - \frac{\epsilon_c}{1+\epsilon_c} (\vec{R}_{gc}\cdot\vec{e}_c)\vec{e}_c, \\
     \label{dOmega_g_two_comps_cs}
     & \dot{\vec{\Omega}}_g + [\vec{\Omega}_c\times \vec{\Omega}_g]= \vec{R}_{cg},
  \end{align}    
  where the following notations have been introduced: 
  $\vec{R}_{ij} = \vec{N}_{ij}/I_j$ and $\vec{\omega}_\mathrm{sf} = \vec{L}_\mathrm{sf}/I_c$.
  Note that $\vec{\omega}_\mathrm{sf}$ is just a convenient notation. In the general case, it is not have the meaning of superfluid angular velocity.
  
  This system of equations has a fixed point corresponding to the uniform rotation of both components with the same angular velocity $\vec{\Omega}_0$  parallel to $\vec{e}_c$.
  In order to consider the linear perturbations to the equilibrium state, it is convenient to introduce small departure vectors:
  \begin{equation}
    \label{eq:mu_two_coms_cs}
    \vec{\mu}_i = \vec{\Omega}_i - \vec{\Omega}_0.
  \end{equation}
  Let us also assume that, with linear in $\vec{\mu_{i}}$ accuracy, the interaction torques {have a simple frictional form}:
  \begin{equation}
    \label{eq:Rij_two_comps}
    \vec{R}_{ij} = \alpha_{ij} (\vec{\mu}_i - \vec{\mu}_j),
  \end{equation}
  { where coefficients $\alpha_{ij} >0 $.}
  From the angular momentum conservation it follows that
  \begin{equation}
    \alpha_{ij} = \frac{I_i}{I_j}\alpha_{ji}.
  \end{equation}

  Substituting expressions \eqref{eq:mu_two_coms_cs} and \eqref{eq:Rij_two_comps} into equations \eqref{dOmega_c_two_comps_cs}, \eqref{dOmega_g_two_comps_cs} and neglecting the quadratically small terms, we obtain
  \begin{align}
       \label{eq:dmu_c_par_two_comps_cs}  
       & \dot{{\mu}}_c^\parallel  - \frac{\alpha_{gc}}{1+\epsilon_c} (\mu_g^\parallel-\mu_c^\parallel) =  0, \\
       \label{eq:dmu_g_par_two_comps_cs}       
       & \dot{{\mu}}_{g}^\parallel - \alpha_{cg} \left( \mu_c^\parallel -\mu_g^\parallel \right)=0, \\
       \label{eq:dmu_c_perp_two_comps_cs}
       & \dot{\mu}_c^\perp - i\tilde{\epsilon}_c \Omega_0 \mu_c^\perp - \alpha_{gc} (\mu_g^\perp - \mu_c^\perp) = 0, \\
       \label{eq:dmu_g_perp_two_comps_cs}
       &\dot{\mu}_g^\perp + i \Omega_0 (\mu_g^\perp - \mu_c^\perp)  - \alpha_{cg}\left(\mu_c^\perp - \mu_g^\perp\right) = 0,
  \end{align} 
  where 
  \begin{equation}
    \label{eq:tilde_eps_two-comps}
    \tilde{\epsilon}_c = \epsilon_c + \frac{\omega_\mathrm{sf}^\parallel}{\Omega_0}
  \end{equation}
  is the c-component effective oblateness parameter.
  One can see that the system of equations falls into two independent parts. The ``parallel'' part describes the evolution of the difference in absolute values of the angular velocities while the ``perpendicular'' part {works} when  the angular velocities decline from the symmetry axis. 
  Substituting $\mu_i^\parallel \propto \exp(p t)$ into equations \eqref{eq:dmu_c_par_two_comps_cs}--\eqref{eq:dmu_g_par_two_comps_cs},
  we obtain 
  \begin{equation}
    \label{eq:par_mode_two_comps_cs}
    p_\parallel = -\alpha _{cg} - \frac{\alpha_{gc}}{1+{\epsilon}_{c}}.
  \end{equation}
  The angular velocity difference, being excited,  decays exponentially at the characteristic time-scale  $\sim p_\parallel^{-1}$.
  
  Since our main goal is  studying the precession, the ``perpendicular" modes are more interesting for us.
  Substituting $\mu_i^\perp \propto \exp(p t)$ into equations \eqref{eq:dmu_c_perp_two_comps_cs}-\eqref{eq:dmu_g_perp_two_comps_cs},
  we obtain the following characteristic equation:
  \begin{equation}
    \label{eq:char_eq_two_comps_cs}
     p^2 + p (i\Omega_0-i\tilde{\epsilon}_c\Omega_0+\alpha_{cg}+\alpha_{gc}) +  \tilde{\epsilon}_c\Omega_0^2-i\tilde{\epsilon}_c\Omega_0\alpha_{cg} = 0.
  \end{equation}
  This equation can be solved exactly. However, it would be more informative to obtain the approximate roots corresponding to different limiting cases. 
  
  Introducing an {auxiliary} interaction parameter,
  \begin{equation}
    \label{eq:sigma}
    \sigma = \frac{\alpha_{cg} + \alpha_{gc}}{\Omega_0},
  \end{equation}
  we can rewrite equation \eqref{eq:char_eq_two_comps_cs} in the following form:
  \begin{equation}
    \label{eq:char_eq_two_comps_cs_wsigm}
    p^2 + i p (1-\tilde{\epsilon}_c)\Omega_0 +  \tilde{\epsilon}_c\Omega_0^2 +  \sigma \Omega_0 (p  -i \tilde{\epsilon}\Omega_0)   = 0,
  \end{equation}
  where 
  \begin{equation}
    \tilde{\epsilon} = \frac{I_c \tilde{\epsilon}_c}{I_c + I_g}
  \end{equation}
  is the effective oblateness parameter of the star as a whole.
  
  If the interaction between the components is weak ($\sigma\ll1$),  the roots of equation \eqref{eq:char_eq_two_comps_cs_wsigm} can be found in the form of the expansion in $\sigma$.
  Keeping the linear in $\sigma$ terms, after returning to the initial interaction coefficients we obtain \citep{SedrakianWassermanCordes1999}:
  \begin{align}
    \label{eq:rot_mode_two_comps_cs}
    &p_1 \approx - i \Omega_0 -\alpha_{cg} -\frac{\alpha_{gc}}{1+\tilde{\epsilon}_c}, \\
    \label{eq:prec_mode_two_comps_cs}
    &p_2 \approx i \tilde{\epsilon}_c \Omega_0 - \frac{\tilde{\epsilon}_c}{1+\tilde{\epsilon}_c}\alpha_{gc}.
  \end{align}
  Here, $p_1$ is the rotational mode corresponding to the angular velocities misalignment.
  According to expression \eqref{eq:rot_mode_two_comps_cs} the angular velocity difference vector ,  $\vec{\nu}_{cg} =\vec{\Omega}_c-\vec{\Omega}_g$, rotates with angular velocity $-\vec{\Omega}_0$ if it is observed from the co-rotating frame of reference. 
  Therefore, in the inertial frame of reference, the direction of vector $\vec{\nu}_{cg}$ is constant.
  The absolute value of vector $\vec{\nu}_{cg}$ decreases with time.
  The characteristic time-scale of the mode decay is
  \begin{equation}
    \label{eq:tau_d_rot_mode_two_comps_cs}
    \tau_d \sim   \left(\alpha_{cg} +\frac{\alpha_{gc}}{1+\tilde{\epsilon}_c}\right)^{-1} \approx \frac{1}{\sigma\Omega_0} \gg \frac{1}{\Omega_0}.
  \end{equation}
  Note, that, in contrast to ``parallel" mode $p_\parallel$, this one contains $\tilde{\epsilon}_{c}$ instead of $\epsilon_{c}$.

  The second mode represents the free precession motion. The imaginary part of  mode \eqref{eq:prec_mode_two_comps_cs} is similar to  mode \eqref{eq:prec_mode_rigid_star_wp}.
  But now it has a real part and, hence, it decays with time as well as the rotational mode.
  The characteristic time-scale of the precession damping is 
  \begin{equation}
    \label{eq:tau_d_prec_mode_two_comps_cs}
    \tau_d \sim {\tilde{\epsilon}_c\alpha_{gc}} \gg \frac{1}{\tilde{\epsilon}_c\Omega_0}.
  \end{equation}  
  {The star will complete of the order of $\Omega_0/\alpha_{gc}\gg1$ precession cycles before the angular velocities align with the symmetry axis but the precession, being, as in the rigid-body case, governed by the pinned superfluid, remains fast.}


  In the opposite limiting case of strong interaction ($\sigma\gg1$), 
  the approximate roots obtained as the expansion in $\sigma^{-1}$ are \citep{SedrakianWassermanCordes1999}
  \begin{align}
    \label{eq:rot_mode_two_comps_cs_si}
    &p_1 \approx -i  \left( 1 + \tilde{\epsilon} -\tilde{\epsilon}_c\right)\Omega_0 - \alpha_{gc} - \alpha_{cg}, \\
    \label{eq:prec_mode_two_comps_cs_si}
    &p_2 \approx i \tilde{\epsilon}\Omega_0 - \frac{{(1+\tilde{\epsilon})(\tilde{\epsilon}_c - \tilde{\epsilon})}}{\alpha_{cg}+\alpha_{gc}}\Omega_0^2.
  \end{align}
  Again, we have one rotational mode, $p_1$, and one precession mode, $p_2$. 
  The real part of mode \eqref{eq:rot_mode_two_comps_cs_si} is much larger than its imaginary part. Therefore, being excited, this mode dacays before the first rotational cycle completes. Hence, the rotational mode is not oscillatory in the strong interaction limit.
  
  The real part of \eqref{eq:prec_mode_two_comps_cs_si} relates to the imaginary part approximately as $(\alpha_{ij}/\Omega_0)^{-1} \ll 1$. 
  Therefore, the precession mode is slowly damped as well as the weak-interaction precession mode.
  The imaginary part of mode \eqref{eq:prec_mode_two_comps_cs_si} relates to the imaginary part of mode \eqref{eq:prec_mode_two_comps_cs} as $I_c/(I_c+I_g)$.
  {Hence, the period of precession is longer in the strong-interaction limit but the precession is still governed by the pinned superfluid.
  Passing $(\alpha_{ij}/\Omega_0)$ to infinity we reproduce the rigid-body precession mode \eqref{eq:prec_mode_rigid_star_wp}.
  
   In sum, the introduction of the internal dissipation in the way described above leads to the damping of the precession but the precession frequency remains high. However, the two-component approximation gives us another possible configuration.}
  

\section{Two-component system with pinning in the core}
  \label{sec:two_comps_gs}
  {Let us {shift} the pinned superfluid into the inner g-component.
  With all the other assumptions being kept, the angular momenta of the components would have the following form: }
  \begin{align}
    \label{eq:Mc_two_comp_gs}
    &\vec{M}_c = I_c \vec{\Omega}_c + I_c \epsilon_c (\vec{\Omega}_c\cdot\vec{e}_c)\vec{e}_c, \\
    \label{eq:Mg_two_comp_gs}
    &\vec{M}_g = {I}_g \vec{\Omega}_g  +  \vec{L}_\mathrm{sf}.
  \end{align}
  
  Substituting expressions \eqref{eq:Mc_two_comp_gs} and \eqref{eq:Mg_two_comp_gs} into equations \eqref{eq:dMc_two_comp} and \eqref{eq:dMg_two_comp}, and changing to the c-component frame of reference, we obtain the equations for the angular velocities: 
  \begin{align}
    \label{eq:dOmega_c_two_comps_gs}
    &\dot{\vec{\Omega}}_c  + {\epsilon}_c(\vec\Omega_c\cdot \vec{e}_c)  [\vec{\Omega} \times \vec{e}_c ] = \vec{R}_{gc}  - \frac{\epsilon_c}{1+\epsilon_c} (\vec{R}_{gc}\cdot\vec{e}_c)\vec{e}_c,  \\
    \label{eq:dOmega_g_two_comps_gs}
    &\dot{\vec{\Omega}}_g + [\vec{\Omega}_c\times \vec{\Omega}_g] + [\vec{\Omega}_g\times\vec{\omega}_\mathrm{sf}] = \vec{R}_{cg},
  \end{align}  
  where vector
  \begin{equation}
    \label{eq:omega_sf}
    \vec{\omega}_\mathrm{sf} = \frac{\vec{L}_\mathrm{sf}}{I_g}
  \end{equation}  
  has been reintroduced.
  
  Since the angular momentum of the pinned superfluid  is now fixed in the g-component, for an  observer in the c-component we have the following equation for vector $\vec{\omega}_\mathrm{sf}$:
  \begin{equation}
    \label{eq:dom_sf_two_comp_gs}
    \dot{\vec{\omega}}_\mathrm{sf} = (\vec{\Omega}_g - \vec{\Omega}_c) \times\vec{\omega}_\mathrm{sf}.
  \end{equation}  
  
  The rotation of the star  is stationary if 
  $\vec{\Omega}_c\parallel\vec{e}_c$, $\vec{\Omega}_g\parallel\vec{\omega}_\mathrm{sf}$ and $\vec{\Omega}_c=\vec{\Omega}_g$.  
  Considering the linear perturbations to this equilibrium state, let us first multiply equation \eqref{eq:dom_sf_two_comp_gs} by vector $\vec{e}_c$ scalarly.
  In terms of ``parallel'' and ``perpendicular'' components the result has the form
  \begin{equation}
    \label{eq:dom_sf_par_two_comps_gs_tmp}
    \dot{\omega}_\mathrm{sf}^\parallel =  -\Re \left[i \omega_\mathrm{sf}^\perp (\mu_g^\perp - \mu_c^\perp)^\dagger\right],
  \end{equation}
  where $\dagger$ denotes the complex conjugation.
  Since, in the zeroth approximation, vector $\vec{\omega}_\mathrm{sf}$ is parallel to $\vec{e}_c$, quantity ${\omega}_\mathrm{sf}^\perp$ can be treated as a small variable together with four variables  $\mu_c^\parallel$, $\mu_c^\perp$, $\mu_g^\parallel$ and $\mu_g^\perp$, which were introduced earlier.
  Therefore, 
  the right-hand side of equation \eqref{eq:dom_sf_par_two_comps_gs_tmp} is quadratically small. Thus, operating in the linear approximation we can put
  \begin{equation}
    \label{eq:dom_sf_par_two_comps_gs}
     {\omega}_\mathrm{sf}^\parallel = \const.    
  \end{equation}
  Keeping this in mind, we can formulate a  system of equations for the linear perturbations:
  \begin{align}
    \label{eq:dmu_c_par_two_comps_gs}
    &\dot{\mu}_c^\parallel  - \frac{\alpha_{gc}}{1+\epsilon_c} (\mu_g^\parallel-\mu_c^\parallel) =  0 \\
    \label{eq:dmu_g_par_two_comps_gs}
    &\dot{\mu}_{g}^\parallel - \alpha_{cg} ( \mu_c^\parallel -\mu_g^\parallel )=0, \\
    \label{eq:dmu_c_perp_two_comps_gs}
    &\dot{\mu}_c^\perp - i{\epsilon}_c \Omega_0 \mu_c^\perp - \alpha_{gc} (\mu_g^\perp - \mu_c^\perp) = 0, \\
    \nonumber
    &\dot{\mu}_g^\perp  - i \Omega_0 \mu_c^\perp + i  \Omega_0 (1-\tilde{\epsilon}_g) \mu_g^\perp   + i\tilde{\epsilon}_g \Omega_0^2 \frac{\omega_\mathrm{sf}^\perp}{\omega_\mathrm{sf}^\parallel}  \\
    \label{eq:dmu_g_pepr_two_comps_gs}
    & \ \ \ \ \ \ \ \ \ \ \ \ \ \ \ \ \ \ \ \ \ \ \ \ \ \ \ \ \ \  \ \ \ \ \ \ \ \ \ \ \  - \alpha_{cg}(\mu_c^\perp - \mu_g^\perp) = 0, \\
    \label{eq:dom_sf_perp_two_comps_gs}
    &\dot{\omega}_\mathrm{sf}^\perp + i  \omega_\mathrm{sf}^\parallel (\mu_g^\perp-\mu_c^\perp) = 0.
  \end{align}
  Here, we have introduced the g-component effective oblateness parameter as $\tilde{\epsilon}_g = \omega_\mathrm{sf}^\parallel/\Omega_0$.
  Note that due to equality \eqref{eq:dom_sf_par_two_comps_gs} the system of equations falls into two independent subsystems.
  
  Since equations \eqref{eq:dmu_c_par_two_comps_gs} and \eqref{eq:dmu_g_par_two_comps_gs} coincide with equations \eqref{eq:dmu_c_par_two_comps_cs} and \eqref{eq:dmu_g_par_two_comps_cs}, the ``parallel" mode remains the same.
  Let us consider  the  ``perpendicular'' modes.
  Substituting $\mu_i^\perp, \omega_g^\perp \propto \exp(p t)$  into equations \eqref{eq:dmu_c_perp_two_comps_gs}--\eqref{eq:dom_sf_perp_two_comps_gs} 
  we obtain the third-order characteristic equation
  \begin{align}
     \nonumber
    &p^3 + i p^2 (1- {\epsilon}_c-\tilde{\epsilon}_g)\Omega_0 +  p ({\epsilon}_c + \tilde{\epsilon}_g - {\epsilon}_c\tilde{\epsilon}_g)\Omega_0^2  \\
     &\ \ \ \ \ \ \ \ \ \ \ \ \ \ \ \ \ \ \ \ \ \ \ \ \  -i \epsilon_c\tilde{\epsilon}_g\Omega_0^3  + \sigma \Omega_0(p^2   - i p \tilde{\epsilon} \Omega_0) = 0,
     \label{eq:char_eq_two_comps_gs}
  \end{align}
  where we have again used the interaction parameter \eqref{eq:sigma} and introduced the whole star oblateness parameter as
  \begin{equation}
    \tilde{\epsilon} = \frac{I_c {\epsilon}_c + I_g \tilde{\epsilon}_g }{I_c + I_g}.
  \end{equation}

  
  Assuming  weak interaction  between the stellar components ($\sigma \ll 1$) we obtain
  \begin{align}
    \label{eq:rot_mode_two_comps_gs}
    &p_1 \approx - i \Omega_0 -\frac{\alpha_{cg}}{1+\tilde{\epsilon}_g} -\frac{\alpha_{gc}}{1+{\epsilon}_c}, \\
    \label{eq:prec_c_mode_two_comps_gs}
    &p_2 \approx i {\epsilon}_c \Omega_0 - \frac{{\epsilon}_c}{1+{\epsilon}_c}\alpha_{gc}, \\
    \label{eq:prec_g_mode_two_comps_gs}
    &p_3 \approx i \tilde{\epsilon}_g \Omega_0 - \frac{\tilde{\epsilon}_g}{1+\tilde{\epsilon}_g}\alpha_{cg}.
  \end{align}  
  The first mode is the rotational mode, similar to mode \eqref{eq:rot_mode_two_comps_cs}.
  The second and the the third ones are the precession modes, each of which represents the precession motion of a single component.
  Mode $p_3$ corresponds to the g-component fast precession caused by the superfluid pinned to it.
  The c-component precession mode $p_2$, in contrast to mode \eqref{eq:prec_mode_two_comps_cs}, contains the real (not ``effective'') oblateness parameter. 
   {  Hence, the c-component precession period is not constrained by the pinned superfluid and therefore  it can be sufficiently long if $\epsilon_c$ is small enough.}
  Strictly speaking, oblateness parameter $\epsilon_c$  can be both positive (oblate star) and negative (prolate star). In the second case the mode is unstable.
  The reason for this is as follows. Introducing  {internal dissipation} we allow the star to convert mechanical energy into thermal energy. 
  However, { the dissipation} cannot substantially affect the stellar angular momentum. Thus, the star tends to the state with minimum possible mechanical energy for a given angular momentum. 
  This state is rotation about the major principal axis \citep{LandauLifshitz_Mech}. If $\epsilon_c>0$, this axis is $\vec{e}_c$. Otherwise, it is an axis perpendicular to $\vec{e}_c$.
  In the second case, one just needs to redefine the basic equilibrium state.
  For definiteness, further we will assume that $\epsilon_c>0$.
  
  In  the case of strong interaction, the approximate roots of equation \eqref{eq:char_eq_two_comps_gs} are the following:
  \begin{align}
    \label{eq:rot_mode_two_comps_gs_si}
    &p_1 \approx -i (1+\tilde{\epsilon}-\epsilon_c-\tilde{\epsilon}_g)\Omega_0 - \alpha_{gc}- \alpha_{cg}, \\
    \label{eq:prec1_mode_two_comps_gs_si}
    &p_2 \approx i \tilde{\epsilon}\Omega_0 - \frac{{(1+\tilde{\epsilon})(\tilde{\epsilon}-{\epsilon}_c)(\tilde{\epsilon}_g - \tilde{\epsilon})}}{\tilde{\epsilon}(\alpha_{cg}+\alpha_{gc})}\Omega_0^2\\
    &p_3 \approx i \frac{{\epsilon}_c\tilde{\epsilon}_g}{\tilde{\epsilon}} \frac{\tilde{\epsilon}({\epsilon}_c+\tilde{\epsilon}_g)-{\epsilon}_c\tilde{\epsilon}_g(1+\tilde{\epsilon})}{\tilde{\epsilon}^2(\alpha_{cg}+\alpha_{gc})^2}\Omega_0 -\frac{{\epsilon}_c\tilde{\epsilon}_g}{\tilde{\epsilon}}\frac{\Omega_0^2}{\alpha_{cg}+\alpha_{gc}}.
  \end{align}
  The first two of these are similar to the corresponding roots obtained for the c-component superfluid model (cf. \eqref{eq:rot_mode_two_comps_cs_si} and \eqref{eq:prec_mode_two_comps_cs_si}). It is easy to verify that combination $(\tilde{\epsilon}-{\epsilon}_c)(\tilde{\epsilon}_g - \tilde{\epsilon})$ is always positive and hence the real part of mode \eqref{eq:prec1_mode_two_comps_gs_si} is always negative. Therefore,  mode \eqref{eq:prec1_mode_two_comps_gs_si} is damped.
  The third mode arises when both $\epsilon_c$ and $\tilde{\epsilon}_g$ are non-zero. However, it is easy to see that this mode is not oscillatory. 
  
  Thus, if we assume that the pinning takes place in the internal component, the slow long-lived precession  mode can exist in the case of  weak interaction. 
  {It  is also necessary to make sure that this mode is observable.
  Substituting mode \eqref{eq:prec_c_mode_two_comps_gs} into equations \eqref{eq:dmu_c_perp_two_comps_gs} - \eqref{eq:dom_sf_perp_two_comps_gs}, we can express 
  \begin{align}
    \mu_c^\perp &= \big((1+\epsilon_c)\Omega_0 + O(\sigma) \big) \frac{\omega_\mathrm{sf}^\perp}{\omega_\mathrm{sf}^\parallel}, \\
    \mu_g^\perp &= \big( \Omega_0 + O(\sigma) \big) \frac{\omega_\mathrm{sf}^\perp}{\omega_\mathrm{sf}^\parallel},
  \end{align}
  where $O(\sigma)$ are small corrections due to the component interaction.
  Using expressions \eqref{eq:Mc_two_comp_gs} and \eqref{eq:Mg_two_comp_gs} it can also be shown that
  \begin{equation}
    {M}^\perp = \left(M^\parallel  + O(\sigma)\right) \frac{\omega_\mathrm{sf}^\perp}{\omega_\mathrm{sf}^\parallel},
  \end{equation}
  where $\vec{M} = \vec{M}_c + \vec{M}_g$ is the total angular momentum.
  We see that, if the $p_2$ mode is excited, vectors $\vec{\Omega}_g$ and $\vec{L}_\mathrm{sf}$ remain  directed almost along the (fixed in the inertial frame of reference) stellar angular momentum $\vec{M}$ while vectors $\vec{e}_c$ and $\vec{\Omega}_c$ are declined from it. 
  In the case of an infinitely small parameter $\sigma$, this mode corresponds to the independent precession of the external c-component.
   }

\section{Three-component system}
  \label{sec:three_comps}
  Having considered the basic idea, we turn to the full model.   
  Let us assume that the neutron star consists of three dynamically distinguished components that we will denote as c-, g-, and r-component.
  
  The c-component is the outer component.
  {The c-component as before is assumed to have axisymmetric shape.}
  Its angular momentum can be represented as 
  \begin{equation}
    \label{eq:Mc_three_comps}
    \vec{M}_c = I_c \vec{\Omega}_c + I_c \epsilon_c (\vec{\Omega}_c\cdot\vec{e}_c)\vec{e}_c.
  \end{equation}
  
  The g-component is an inner component. 
  {It consists of  a ``normal" fraction and a pinned superfluid fraction.
  The ``normal" fraction, for more generality, is assumed to have axisymmetric shape. 
  It is characterized by moment of inertia $I_g$ and oblateness parameter $\epsilon_g$. 
  The ``normal" fraction rotates with angular velocity $\vec{\Omega}_g$.
  The pinned superfluid contains angular momentum $\vec{L}_\mathrm{sf}$. 
  Thus, the total angular momentum of the g-component can be represented as }
  \begin{equation}
    \label{eq:Mg_three_comps}
    \vec{M}_{g} = I_g\vec{\Omega}_g + I_g \epsilon_g (\vec{\Omega}_g\cdot\vec{e}_g)\vec{e}_g + \vec{L}_\mathrm{sf},
  \end{equation}
  where $\vec{e}_g$ is unit vector directed along the g-component symmetry axis.
We will assume that vector $\vec{L}_\mathrm{sf}$ is directed along the g-component symmetry axis, i.e.
  \begin{equation}
    \label{eq:Lg_assumption}
    \vec{e}_g = \frac{\vec{L}_\mathrm{sf}}{{L}_\mathrm{sf}} = \frac{\vec{\omega}_\mathrm{sf}}{\omega_\mathrm{sf}},
  \end{equation}
  where $\vec{\omega}_\mathrm{sf}$ is defined by expression \eqref{eq:omega_sf}.
  
  The r-component is the other inner component. It 
  rotates with angular velocity $\vec{\Omega}_{r}$.
  Its possible asymmetries would not lead to substantially new effects but would complicate the expressions.
  Hence, we will assume this component to be spherically symmetric.
  Therefore, the angular momentum of the r-component is assumed to have the form
  \begin{equation}
   \label{eq:Mr_three_comps}
   \vec{M}_r = {I}_r \vec{\Omega}_r.
  \end{equation}
 
  The angular momentum conservation law can be represented in the form of three equations:
  \begin{equation}
    \label{eq:dMi_three_comps}
    d_t\vec{M}_i = \sum_{j \neq i} \vec{N}_{ji},
  \end{equation}
  where $i, \ j = c, \ g, \ r$ and $\vec{N}_{ij}$ is the torque acting on the $i$th component from the $j$th component.
  In the general case, all the components are supposed to interact with each other.
  
  The further analysis is similar to that performed for the two-component system.
  We will  describe it briefly mentioning the main steps and focusing only on new features.
  Substituting expressions \eqref{eq:Mc_three_comps}-\eqref{eq:Mr_three_comps} into equations \eqref{eq:dMi_three_comps} and changing to the frame  of reference co-rotating with the c-component we obtain the following equations for the angular velocities:
  \begin{align}
    &\dot{\vec{\Omega}}_c  + \epsilon_c (\dot{\vec{\Omega}}_c\cdot \vec{e}_c) \vec{e}_c + {\epsilon}_c(\vec\Omega_c\cdot \vec{e}_c)  [\vec{\Omega} \times \vec{e}_c ] = \vec{R}_{gc} + \vec{R}_{rc},  
    \label{eq:dOmega_c_three_comps}    \\
    \nonumber
    &\dot{\vec{\Omega}}_g + \epsilon_g \left(\dot{\vec{\Omega}}_g\cdot \frac{\vec{\omega}_\mathrm{sf}}{\omega_\mathrm{sf}}\right) \frac{\vec{\omega}_\mathrm{sf}}{\omega_\mathrm{sf}}+ [\vec{\Omega}_c\times \vec{\Omega}_g]  \\
    \label{eq:dOmega_g_three_comps}
    & \  +\left(\epsilon_g \left(\vec{\Omega}_g\cdot \frac{\vec{\omega}_\mathrm{sf}}{\omega_\mathrm{sf}}\right) + \omega_\mathrm{sf} \right) \left[\vec{\Omega}_g\times\frac{\vec{\omega}_\mathrm{sf}}{\omega_\mathrm{sf}}\right] = \vec{R}_{cg} + \vec{R}_{rg}, \\
    \label{eq:dOmega_r_three_comps}
    &\dot{\vec{\Omega}}_r + [\vec{\Omega}_c\times \vec{\Omega}_r]  = \vec{R}_{cr} + \vec{R}_{gr}.
  \end{align}
  From the angular momentum conservation it is follows that 
  \begin{equation}
    \label{eq:R_ij_relation}
    \vec{R}_{ij} = - \frac{I_i}{I_j}\vec{R}_{ji}.
  \end{equation}
  In order to close this system, equation \eqref{eq:dom_sf_two_comp_gs} should be added. 
  
  { It is assumed that if}
  \begin{equation}
   \label{eq:unif_rot}
   \vec{\Omega}_c = \vec{\Omega}_g = \vec{\Omega}_r.
  \end{equation}
  and $\vec{\Omega}_c\parallel\vec{\omega}_\mathrm{sf}\parallel\vec{e}_c$, the system is in stable equilibrium.
  The reasoning leading to approximate equality \eqref{eq:dom_sf_par_two_comps_gs} remains valid. Hence,  if the star rotates near the fixed point,
  there are 
  seven small variables, namely $\mu_c^\parallel$, $\mu_c^\perp$, $\mu_g^\parallel$, $\mu_g^\perp$, $\mu_r^\parallel$, $\mu_r^\perp$ and $\omega_\mathrm{sf}^\perp$.

  We assume
  that
  there are no preferred directions for the components interaction except the angular velocities. However, since  the exact interaction mechanisms are not specified at this point, we consider a more general form of vectors $\vec{R}_{ij}$.
  Namely,
  \begin{align}
    \label{eq:R_tree_comps}
    &{R}_{ij}^\parallel = \alpha_{ij} ({\mu}^{\parallel}_{i} - {\mu}^{\parallel}_{j}), \\
    \label{eq:R_tree_comps_perp}
    &{R}_{ij}^\perp = \xi_{ij} ({\mu}^{\perp}_{i} - {\mu}^{\perp}_{j}),
  \end{align}
  where $\xi_{ij} = \beta_{ij} + i \gamma_{ij}$ and $\alpha_{ij}$, $\beta_{ij}$, $\gamma_{ij}$ are the phenomenological interaction constants. 
  The particular case of simple linear friction corresponds to $\beta_{ij} = \alpha_{ij}, \gamma_{ij} = 0$.
  If we consider, for instance, the mutual friction interaction between the $i$- and $j$-component, one of which is superfluid, the coefficients can be represented in the following form \citep{SedrakianWassermanCordes1999,BGT2013}:
  \begin{equation}
    \alpha_{ij} = 2 \Omega_0 \frac{x}{1+x^2}, \  \beta_{ij} = \Omega_0 \frac{x}{1+x^2}, \  \gamma_{ij} = - \Omega_0 \frac{x^2}{1+x^2}
  \end{equation}
  where $x$ is the coupling parameter. 
  From relations \eqref{eq:R_ij_relation} it follows that
  \begin{equation}
    \label{eq:coeffs_relations_three_comps}
    \alpha_{ij} = \frac{I_i}{I_j}
     \alpha_{ji}, \ \ \ \beta_{ij} = \frac{I_i}{I_j} \beta_{ji}, \ \ \ \gamma_{ij} = \frac{I_i}{I_j} \gamma_{ji}.
  \end{equation}

  The linearized system of equations has the form:
  \begin{align}
    \label{eq:dmu_c_par_three_comps}
    &(1+\epsilon_c)\dot{\mu}_c^\parallel + \left(\alpha_{gc} + \alpha_{rc} \right) \mu_c^\parallel - \alpha_{gc} \mu_g^\parallel- \alpha_{rc} \mu_r^\parallel =  0 \\
    \label{eq:dmu_g_par_three_comps}
    &(1+\epsilon_g)\dot{\mu}_g^\parallel + \left(\alpha_{cg} + \alpha_{rg} \right) \mu_g^\parallel - \alpha_{cg} \mu_c^\parallel- \alpha_{rg} \mu_r^\parallel =  0 \\
    \label{eq:dmu_r_par_three_comps}
    &\dot{\mu}_r^\parallel + \left(\alpha_{cr} + \alpha_{gr} \right) \mu_r^\parallel - \alpha_{cr} \mu_c^\parallel- \alpha_{gr} \mu_g^\parallel =  0 \\
    \label{eq:dmu_c_perp_three_comps}
    &\dot{\mu}_c^\perp + \left( \xi_{gc} + \xi_{rc} - i{\epsilon}_c \Omega_0 \right) \mu_c^\perp - \xi_{gc} \mu_g^\perp - \xi_{rc} \mu_r^\perp = 0, \\
    \nonumber
    &\dot{\mu}_g^\perp  + \left(\xi_{cg} + \xi_{rg}  + i  \Omega_0 (1-\tilde{\epsilon}_g) \right)\mu_g^\perp  - \left( i \Omega_0   + \xi_{cg}\right)\mu_c^\perp      \\
    \label{eq:dmu_g_pepr_three_comps}
    & \ \ \ \ \ \ \ \ \ \ \ \ \ \ \ \ \ \ \ \ \ \ \ \ \ \ \ \ \ \  \ \ \ \  - \xi_{rg}\mu_r^\perp + i\tilde{\epsilon}_g \Omega_0^2 \frac{\omega_g^\perp}{\omega_g^\parallel} = 0, \\
    \nonumber
    &\dot{\mu}_r^\perp  + \left(\xi_{cr} + \xi_{gr}  + i  \Omega_0  \right)\mu_r^\perp  - \left( i \Omega_0   + \xi_{cr}\right)\mu_c^\perp       \\
    \label{eq:dmu_g_pepr_three_comps}
    & \ \ \ \ \ \ \ \ \ \ \ \ \ \ \ \ \ \ \ \ \ -\xi_{gr}\mu_g^\perp  = 0,  \\
    \label{eq:dom_sf_perp_three_comps}
    &\dot{\omega}_\mathrm{sf}^\perp + i  \omega_\mathrm{sf}^\parallel (\mu_g^\perp-\mu_c^\perp) = 0,
  \end{align}
  { where the g-component effective oblateness parameter is equal to
  \begin{equation}
    \label{eq:epsilon_g_three_comps}
    \tilde{\epsilon}_g = \epsilon_g + \frac{L_\mathrm{sf}^\parallel}{I_g \Omega_0} = \epsilon_g + \frac{\omega_\mathrm{sf}^\parallel}{\Omega_0}.
  \end{equation}  
  }
  As before, the system falls into ``parallel'' and ``perpendicular'' parts.
  
  Let us first consider ``parallel'' modes. Substituting $\mu_i^\parallel \propto \exp (p t)$ into equations \eqref{eq:dmu_c_par_three_comps}--\eqref{eq:dmu_r_par_three_comps} we obtain the following second-order characteristic equation: 
  \begin{align}
    \nonumber
    &p^{2} + \Omega_0(\sigma_{cg} + \sigma_{cr} +\sigma_{gr})p +\Omega_0^2\left(\sigma_{cg}\sigma_{cr} + \sigma_{cg}\sigma_{gr} + \sigma_{cr}\sigma_{gr}\right)  \\
    \label{eq:char_equations_par_modes_three_comps}
    &- \frac{\alpha_{gc}\alpha_{rc}}{(1+\epsilon_c)^2} - \frac{\alpha_{cg}\alpha_{rg}}{(1+\epsilon_g)^2}  - {\alpha_{cr}\alpha_{gr}}=0.
  \end{align}
  Here, we have introduced three interaction parameters 
  \begin{equation}
    \sigma_{ij} = \frac{1}{\Omega_{0}}\left(\frac{\alpha_{ij}}{1+\epsilon_j} + \frac{\alpha_{ji}}{1+\epsilon_i}\right),
  \end{equation}
  it is  easy to see that $\sigma_{ij} = \sigma_{ji}$.
  The exact roots of equation \eqref{eq:char_equations_par_modes_three_comps} can be given.
  However, instead of that we focus on the one particular case of 
  \begin{equation}
    \label{eq:sigma_cond} 
    \sigma_{cg}\gg\sigma_{cr}, \ \sigma_{gr}
  \end{equation}
  which appears to be the most suitable for the glitch model considered in Section \ref{sec:glitch}.
   With relations \eqref{eq:coeffs_relations_three_comps} the roots can be represented in the form 
  \begin{align}
    \nonumber
    p_{+} &\approx   - \sigma_{cg}\Omega_0 - \frac{(1+\epsilon_g)I_g \alpha_{rc} + (1+\epsilon_c)I_c \alpha_{rg}}{ (1+\epsilon_g)I_g+(1+\epsilon_c)I_c}  \\
    \label{eq:par_mode_+_three_comps}
    &\approx -\frac{\alpha_{cg}}{1+\epsilon_g}  - \frac{\alpha_{gc}}{1+\epsilon_c}, \\ 
    \nonumber
        p_{-} &\approx -(\sigma_{cr} + \sigma_{gr})\Omega_0 + \frac{(1+\epsilon_g)I_g \alpha_{rc} + (1+\epsilon_c)I_c \alpha_{rg}}{ (1+\epsilon_g)I_g+(1+\epsilon_c)I_c}  \\
        \label{eq:par_mode_-_three_comps}
       &= - \frac{(1+\epsilon_c)I_c + (1+\epsilon_g)I_g + I_r}{(1+\epsilon_c)I_c + (1+\epsilon_g)I_g}(\alpha_{cr} + \alpha_{gr}).
  \end{align}
  In this particular case, $p_+\gg p_-$. 
  
  Let us turn to the ``perpendicular" part.
  We do not give here the fourth-order characteristic equation  because of its awkwardness.
%
  By analogy with the ``parallel" case we can introduce three interaction parameters 
  \begin{equation}
    Z_{ij} = \frac{\xi_{ij} + \xi_{ji}}{\Omega_0}.
  \end{equation}
  In the weak-interaction limit ($|Z_{ij}|\ll1$) the approximate expressions for the modes can be represented in the form:
  \begin{align}
    \label{eq:rot_mode_1_three_comps}
    &p_1 \approx -i \Omega_0 +\delta p_1 \\
    \label{eq:rot_mode_2_three_comps}
    &p_2 \approx  -i \Omega_0 + \delta p_2 \\
    \label{eq:prec_mode_c_three_comps}
    &p_3 \approx  i \epsilon_c \Omega_0 - \frac{\epsilon_c}{1+\epsilon_c}\left(\xi_{gc} + \xi_{rc}\right)\\
    \label{eq:prec_mode_g_three_comps}
    &p_4 \approx  i\tilde{\epsilon}_g \Omega_0 - \frac{\tilde{\epsilon}_g}{1+\tilde{\epsilon}_g}(\xi_{rg} +\xi_{cg}).
  \end{align}
  With the introduction of the third component the rotational mode becomes degenerate. It splits into two modes when the dissipation corrections $\delta p_1$ and $\delta p_2$ are taken into account. They can be found as the roots of the following equation:
  \begin{align}
    \nonumber
    &\delta p^2 - \delta p \left[ (1+\epsilon_c)\xi_{cg} + (1+\tilde{\epsilon}_g)(\xi_{gc}+\xi_{rc}) +\right.\\
    \nonumber
    &\left. + (1+\epsilon_c)\xi_{rg} +(1+\epsilon_c)(1+\tilde{\epsilon}_g)(\xi_{gr}+\xi_{cr}))  \right] - \\
    \nonumber
    &-(1+\epsilon_c)(\xi_{cr}\xi_{rg} + \xi_{cg}\xi_{cr}+ \xi_{cg}\xi_{rc}) - \\
    \nonumber
    &-(1+\tilde{\epsilon}_g)(\xi_{gc}\xi_{cr}+\xi_{gc}\xi_{gr}+\xi_{rc}\xi_{gr})- \\
    \label{eq:dp}
    &-(\xi_{gc}\xi_{rg}+\xi_{rc}\xi_{rg}+\xi_{cg}\xi_{rc}) = 0.
  \end{align}
  To make some estimations by analogy with "parallel" modes we assume that 
  $|Z_{cg}| \gg |Z_{cr}|,  |Z_{gr}|$.
  This allows us to treat relations $Z_{cr}/Z_{cg}$ and $Z_{gr}/Z_{cg}$  as small parameters and represent the roots of equation \eqref{eq:dp} in the form of expansion in them.
  The resulting expressions formally coincide with expressions \eqref{eq:par_mode_+_three_comps} and \eqref{eq:par_mode_-_three_comps} with corresponding replacements of $\alpha_{ij}$ by $\xi_{ij}$ and $\epsilon_g$ by $\tilde{\epsilon}_g$.
  Since our goal is estimating the damping time-scales,
  for the sake of brevity, we neglect all the oblateness parameters, assuming their smallness.
  To the first order in the small interaction parameters we have
  \begin{align}
    &\delta p_1 \approx - \xi_{cg} - \xi_{gc} - \frac{I_g\xi_{rc} + I_c\xi_{rg}}{I_c+I_g}, \\
    &\delta p_2 \approx - \frac{I_c(\xi_{rc}+\xi_{cr}+\xi_{gr}) + I_g(\xi_{cr}+\xi_{gr}+\xi_{rg})}{I_c+I_g}.
  \end{align}
  Another possible case is $\tilde{\epsilon}_g\gg$1, for which we have
  \begin{align}
    &\delta p_1 \approx - \frac{\xi_{cg}+ \tilde{\epsilon}_g\xi_{gc}}{\tilde{\epsilon}_g} - \frac{I_c\xi_{rg} + \tilde{\epsilon}_g^2I_g\xi_{rc}}{\tilde{\epsilon}_g(I_c + \tilde{\epsilon}_g I_g)},\\
    &\delta p_2 \approx - \frac{I_c(\xi_{rc}+\tilde{\epsilon}_g\xi_{cr}+\tilde{\epsilon}_g\xi_{gr}) + I_g(\tilde{\epsilon}_g\xi_{cr}+\tilde{\epsilon}_g\xi_{gr}+\xi_{rg})}{I_c + \tilde{\epsilon}_gI_g}.
  \end{align}
  Thus, assuming that the cg-interaction is much stronger than the other two, we obtain the hierarchy of damping time-scales. Namely, the real part of $p_+$ and $p_1$ is determined by the strength of the strongest interaction (i.e. cg-intraction) while $p_-$ and $p_2$ modes are damped rather due to the cr- and  gr-interactions.

  Mode $p_3$ corresponds to the c-component slow precession. It is almost similar to mode \eqref{eq:prec_c_mode_two_comps_gs} obtained in the framework of the two-component model. However, let us note that the precession period, in the general case, would be equal to 
  \begin{equation}
    \label{eq:Tp_three_comps}
   T_p = \frac{P}{\epsilon_c \left(1 - \frac{\gamma_{gc}  + \gamma_{rc}}{\Omega_0}\right)}.
 \end{equation}
  
  Mode $p_4$ corresponds to the fast precession of the g-component. 
  We do not take into account the possible non-sphericity of the r-component. However,  one can see that in the weak-interaction limit each component precesses almost independently. Hence, the precession mode for the r-component can easily be obtained from $p_3$ or $p_4$ by the corresponding interchange of the indices.
  
  We restrict ourselves to considering only the weak-interaction limit between all pairs of components.
  {We have seen in Section \ref{sec:two_comps_gs} that the strong-interaction limit 
  between the c- and g-component
  does not allow the observed c-component to precess with long period.  
  }
  Hence, the gc-interaction should be weak. 
  On the other hand, we will see in sec. \ref{sec:glitch} that the cg-interaction should be the strongest among all pair interactions. 
  Thus, the only possible case is the weak interaction between all the components. 
  Recall that by weak interaction we mean that 
  \begin{equation}
    \label{eq:weak_limit_three-comps}
     \frac{\alpha_{ij}}{\Omega_0}, \ \frac{\beta_{ij}}{\Omega_0}, \ \frac{\gamma_{ij}}{\Omega_0} \ll 1.
  \end{equation}
 

\section{Quasi-stationary evolution}
  \label{sec:qstat_evolution}
  
  We have seen that the long-period precession mode can arise if we assume that the region of superfluid pinning
  can
  {rotate}
  relative to the crust.
  {This result was obtained for a freely rotating neutron star perturbed from stable equilibrium.}
  Such a formulation of the problem, however, could be not quite realistic. 
  As it was mentioned at the beginning of the paper, strictly speaking, the rotation of neutron stars is not free. Even isolated neutron stars rotate under the action of electromagnetic torque $\vec{K}$ caused by the rotation of the strong magnetic field anchored to the star \citep{DavisGoldstein1970,BeskinGurevichIstomin1983,Melatos2000}.
  The part of this torque that we have denoted by $\vec{K}_2$ can be easily taken into account by redefinition of the stellar moment of inertia tensor (see Section \ref{sec:rigid_body}).
  As for the second part, $\vec{K}_3$,  we have not taken it into account because its effects are negligibly small at precession time-scales. 
  Now we are going to formulate the equations that allow us to consider the evolution of pulsar rotation at the pulsar life time-scales.
  
  We will consider the three-component model introduced in the previous section. Equations \eqref{eq:dOmega_g_three_comps}, \eqref{eq:dOmega_r_three_comps} and \eqref{eq:dom_sf_two_comp_gs} remain the same. Equation \eqref{eq:dOmega_c_three_comps} should be replaced by a slightly modified one that is more convenient for subsequent consideration.
  First, let us re-establish the third-order external torque term $\vec{K}_3$, which we put on the right-hand side of equation \eqref{eq:dOmega_c_three_comps}.
  After simple rearrangement the equation takes the following form:
  \begin{equation}
    \label{eq:dOm_c_quasistat_tmp}
    \dot{\vec{\Omega}}_c  + \epsilon_c (\dot{\vec{\Omega}}_c\cdot \vec{e}_c) \vec{e}_c   =  \vec{R}_{gc} + \vec{R}_{rc} +\vec{S},      
  \end{equation}
  where we have introduced vector
  \begin{equation}
    \vec{S} = \frac{\vec{K}_3}{I_c} -{\epsilon}_c(\vec\Omega_c\cdot \vec{e}_c)  [\vec{\Omega}_c \times \vec{e}_c ].
  \end{equation}
  Multiplying equation \eqref{eq:dOm_c_quasistat_tmp} by $\vec{e}_c$ and substituting the obtained expression for $(\vec{e}_c\cdot\vec{\Omega}_c)$ back into equation \eqref{eq:dOm_c_quasistat_tmp} we get
  \begin{equation}
    \label{eq:dOm_c_quasistat_tmp_2}
    \dot{\vec{\Omega}}_c = \overrightarrow{\mathrm{RHS}} - \frac{\epsilon_c}{1+\epsilon_c} (\overrightarrow{\mathrm{RHS}}\cdot\vec{e}_c)\vec{e}_c,
  \end{equation}
  where by $\overrightarrow{\mathrm{RHS}}$ we mean the right-hand side of equation \eqref{eq:dOm_c_quasistat_tmp}.
  Recall that the c-component oblateness parameter is supposed to be of the order of $10^{-8}$ or smaller. Therefore, the second term on the right-hand side of equation \eqref{eq:dOm_c_quasistat_tmp_2} is negligibly small. Thus, the c-component equation becomes
  \begin{equation}
    \label{eq:dOm_c_quasistat}
    \dot{\vec{\Omega}}_c  =  \vec{R}_{gc} + \vec{R}_{rc} +\vec{S}.
  \end{equation}
  
  The c-component angular velocity vector time derivative can be represented as a sum of two terms:
  \begin{equation}
    \label{eq:dOm_two_parts}
    \dot{\vec{\Omega}}_c =  \dot{\Omega}_c\vec{e}_\Omega + \Omega_c \dot{\vec{e}}_\Omega,
  \end{equation}
  where $\vec{e}_\Omega = \vec{\Omega}_c/\Omega_c$. Here, the first term represents the change of the angular velocity absolute value while the second term arises due to change of its orientation.
  Again it will be convenient to introduce the ``parallel'' and ``perpendicular'' parts of the vectors. 
  However, in contrast to  previous sections, we define  the parts relative to the direction of vector $\vec{\Omega}_c$ instead of $\vec{e}_c$. 
  Let $\vec{e}_x,  \ \vec{e}_y,  \ \vec{e}_z$ be an orthonormal basis oriented such that $\vec{e}_z \parallel \vec{e}_\Omega$ and $\vec{e}_x \parallel \dot{\vec{e}}_\Omega$.
  Having fixed the basis, we can introduce the following notations:
  \begin{equation}
    \label{eq:V_par_perp_quasistat}
     V^{(\parallel)} = (\vec{V}\cdot\vec{e}_z), \ \ \ {V}^{(\perp)}  = (\vec{V}\cdot\vec{e}_x) + i (\vec{V}\cdot\vec{e}_y),
  \end{equation}  
  where $\vec{V}$ is an arbitrary vector.
  Taking into account expression \eqref{eq:dOm_two_parts} and \eqref{eq:V_par_perp_quasistat}, we can represent equation  \eqref{eq:dOm_c_quasistat} as two equations:
  \begin{align} 
    \label{eq:dOm_c_par_quasistat} 
    &\dot{\Omega}_c  =  R_{gc}^{(\parallel)} + R_{rc}^{(\parallel)}  + S^{(\parallel)},\\
    \label{eq:dOm_c_perp_quasistat}
    &\Omega_c \dot{{e}}_{\Omega}  =  
          {R}_{gc}^{(\perp)} + {R}_{rc}^{(\perp)}  + {S}^{(\perp)}.
  \end{align}
  
  The next step is  linearization. 
  Let us introduce departure vectors 
  \begin{equation}
    \label{eq:mu}
    \vec{\nu}_{ij} = \vec{\Omega}_i - \vec{\Omega}_j = \vec{\mu}_i - \vec{\mu}_j
  \end{equation} 
  We will assume that all the components rotate with almost the same angular velocity and the g-component rotates almost about its symmetry axis $\vec{e}_g$. However, the angle between $\vec{\Omega}_c$ and $\vec{e}_c$ is now not necessarily small.
  Multiplying equation \eqref{eq:dom_sf_two_comp_gs} by vector $\vec{e}_\Omega$
  we obtain
  \begin{equation}
    \label{eq:dom_sf_par_quasistat}
      \dot{\omega}_\mathrm{sf}^{(\parallel)} = - \Re \left[ i {\omega}_\mathrm{sf}^{(\perp)}{\nu}_{gc}^{(\perp)\dagger} + {\omega}_\mathrm{sf}^{(\perp)}\dot{{e}}_\Omega\right].
  \end{equation}
  Since, in the zeroth approximation, vector $\vec{\omega}_\mathrm{sf}$ is parallel to $\vec{e}_\Omega$, quantity ${\omega}_\mathrm{sf}^{(\perp)}$ can be treated as a small variable together with the components of the angular velocity departure vectors.
  Hence, the first term on the right-hand side of equation \eqref{eq:dom_sf_par_quasistat} is straightforwardly quadratically small.
  According to equation \eqref{eq:dOm_c_perp_quasistat} time derivative $\dot{{e}}_{\Omega}$ is linearly small. Therefore, the second term on the right-hand side of equation \eqref{eq:dom_sf_par_quasistat} is quadratically small as well. Thus, the parallel part of vector $\vec{\omega}_\mathrm{sf}$ with linear accuracy can be treated as a constant.
  
  The interaction between the components is described by vectors $\vec{R}_{ij}$. It is assumed that in the small departure approximation (${\nu}_{ij}/\Omega_0\ll 1$) they can be represented in the following form:
  \begin{align}
    \label{eq:R_quasistat}
    &{R}_{ij}^{(\parallel)} = \alpha_{ij} {\nu}^{(\parallel)}_{ij}, \\
    &{R}_{ij}^{(\perp)} = \xi_{ij} {\nu}^{(\perp)}_{ij} ,
  \end{align}
  where  
   $\xi_{ij} = \beta_{ij} + i \gamma_{ij}$ and 
  $\alpha_{ij}$, $\beta_{ij}$, $\gamma_{ij}$ are the interaction constants.   
  If $\vec{e}_\Omega\approx\vec{e}_c$, this representation, up to linear terms, reproduces expressions \eqref{eq:R_tree_comps} and \eqref{eq:R_tree_comps_perp}.
  Hence, the interaction constants $\alpha_{ij}$, $\beta_{ij}$, $\gamma_{ij}$ coincide with the coresponding interaction constants introduced in Section \ref{sec:three_comps}.
  
  Substituting definition \eqref{eq:mu} into equations \eqref{eq:dOmega_g_three_comps}, \eqref{eq:dOmega_r_three_comps} and \eqref{eq:dom_sf_two_comp_gs} and neglecting quadratically small terms, we obtain
  \begin{align}
    \label{eq:dnu_gc_par_quasistat}
    & {\dot{\nu}_{gc}^{(\parallel)}}  = 
      \frac{R_{cg}^{(\parallel)}}{1+\epsilon_g} + \frac{R_{rg}^{(\parallel)}}{1+\epsilon_g}
    - R_{gc}^{(\parallel)} -R_{rc}^{(\parallel)} - S^{(\parallel)},
    \\
    \label{eq:dnu_rc_par_quasistat}
    & {\dot{\nu}_{rc}^{(\parallel)}}  = 
      {R_{cr}^{(\parallel)}}+ {R_{gr}^{(\parallel)}}
    - R_{gc}^{(\parallel)} -R_{rc}^{(\parallel)} - S^{(\parallel)},
    \\
    \nonumber
    & {\dot{{\nu}}_{gc}^{(\perp)}} +  
      i(1-\tilde{\epsilon}_g)  {\Omega}_c{\nu}_{gc}^{(\perp)}
    + i \tilde{\epsilon}_g \Omega_c^2  \frac{{\omega}_\mathrm{sf}^{(\perp)}}{\omega_\mathrm{sf}^\parallel} \\
    \label{eq:dnu_gc_perp_quasistat}
    & \ \ \ \ \ \ \ \ \ \ \ \ \ \ \ \ =  {R}_{cg}^{(\perp)} + {R}_{rg}^{(\perp)}
    - {R}_{gc}^{(\perp)} - {R}_{rc}^{(\perp)} - {S}^{(\perp)},
    \\
    \nonumber
    & {\dot{{\nu}}_{rc}^{(\perp)}} +  
       i{\Omega}_c {\nu}_{rc}^{(\perp)}  \\
    \label{eq:dnu_rc_perp_quasistat}
    & \ \ \ \ \ \ \ \ \ \ \ \ \ \ \ \ =  {R}_{cr}^{(\perp)} + {R}_{gr}^{(\perp)}
    - {R}_{gc}^{(\perp)} - {R}_{rc}^{(\perp)} - {S}^{(\perp)},
    \\
    \label{eq:dom_sf_perp_quasistat}
    & {\frac{\dot{{\omega}}_\mathrm{sf}^{(\perp)}}{\omega_\mathrm{sf}^{(\parallel)}}} +   i {\nu}_{gc}^{(\perp)} = 
    - \frac{1}{\Omega_c}
      \left(
         {R}_{gc}^{(\perp)} + {R}_{rc}^{(\perp)} + {S}^{(\perp)} 
      \right),
  \end{align}   
  where
  \begin{equation}
    \label{eq:epsilon_g_quasistat}
    \tilde{\epsilon}_g = \epsilon_g + \frac{L_\mathrm{sf}^{(\parallel)}}{I_g \Omega_c} = \epsilon_g + \frac{\omega_\mathrm{sf}^{(\parallel)}}{\Omega_c}.
  \end{equation}    
  The equation for ${\nu}_{gr}^{(\parallel)}= {\nu}_{gc}^{(\parallel)} - {\nu}_{rc}^{(\parallel)}$ and ${\nu}_{gr}^{(\perp)} = {\nu}_{gc}^{(\perp)} - {\nu}_{rc}^{(\perp)}$ can be obtained from equations \eqref{eq:dnu_gc_par_quasistat} -- \eqref{eq:dnu_rc_perp_quasistat}.
  Equations \eqref{eq:dnu_gc_par_quasistat}-\eqref{eq:dom_sf_perp_quasistat} together with equations \eqref{eq:dOm_c_par_quasistat} and \eqref{eq:dOm_c_perp_quasistat} form a closed system describing the rotation evolution of our three-component neutron star.
  
  We have linearized the system of equations with respect to  $\nu_{gc}^{(\parallel)}$, $\nu_{rc}^{(\parallel)}$, ${\nu}_{gc}^{(\perp)}$, ${\nu}_{rc}^{(\perp)}$ and ${\omega}_\mathrm{sf}^{(\perp)}$. However, equations \eqref{eq:dOm_c_par_quasistat}, \eqref{eq:dOm_c_perp_quasistat} and \eqref{eq:dnu_gc_par_quasistat} - \eqref{eq:dom_sf_perp_quasistat} contain different combinations with vector $\vec{\Omega}_c$ which is not small. In its present form this system of equations is quite difficult to solve.
  Fortunately, the analysis performed in Section \ref{sec:three_comps} gives us a way to simplify this system.
  {
  Let us highlight the following:
  \begin{enumerate}
    \item \label{item:I}The observed quasi-periodic processes that are interpreted as a manifestation of the long-period precession mode have the time-scale $T_p\sim10^2$ d or longer and, hence, $\epsilon_c \sim 10 ^{-8}$ or smaller.
    \item  \label{item:II} If at least 1 percent of the total moment of inertia is contained in the g-component pinned superfluid, according to expression \eqref{eq:epsilon_g_three_comps} the g-component effective oblateness parameter $\tilde{\epsilon}_g$ cannot be smaller that $\sim 10^{-2}$.
    The estimations that will be done in Section \ref{sec:glitch} provide $\tilde{\epsilon}_g \sim 10$.
      \item  \label{item:III} The interactions between the components are assumed to be weak but not too weak.   
      It is  required that  $\alpha_{ij}/\Omega_c, |\xi_{ij}|/\Omega_c \ll 1$ for all pairs of the components.
  Let us also constrain the coefficient values from the other side assuming that
  \begin{equation}
    \label{eq:qusistat_ineq_val}
    \frac{\epsilon_c\Omega_0}{\tilde{\epsilon}_g(\beta_{cg}+\beta_{gc})}\ll1, \ \ \ \ 
    \frac{\epsilon_c\Omega_0}{\mathrm{min}(\alpha_{ij}, \ \beta_{ij})}\ll1.
  \end{equation}
  \end{enumerate}
  From points \ref{item:I} -- \ref{item:III} it is follows that there is  a gap between the long-period precession time-scale ($\sim T_p$) and the time-scales of all the other modes except the long-period precession one.
  In other words, if we introduce the internal relaxation time-scale $\tau_r$ as a smallest real part of  all the other modes except the long-period precession mode, condition $\tau_r \ll T_p$ is assumed to be satisfied. 
  Such a hierarchy of time-scales allows us to consider the behaviour of the system on the different time-scales separately. }
  
  Let us first consider the evolution of the system at the internal relaxation time-scale. 
  At this time-scale, 
  the absolute value of angular velosity $\vec{\Omega}_c$ and its orientation relative to vector $\vec{e}_c$ can change only slightly.
  Therefore, we can treat vector $\vec{\Omega}_c$ as well as vector $\vec{S}$ as constants.
  In this case, equations \eqref{eq:dnu_gc_par_quasistat} -- \eqref{eq:dom_sf_perp_quasistat} become a closed system of linear equations.
  { The system is inhomogeneous.  
  Therefore, the general solution consists of a particular solution to system \eqref{eq:dnu_gc_par_quasistat} -- \eqref{eq:dom_sf_perp_quasistat} and the general solution to this system with zeroth $\vec{S}$. 
  The first can be found by putting all the time derivatives in equations  \eqref{eq:dnu_gc_par_quasistat} -- \eqref{eq:dom_sf_perp_quasistat} equal to zero. 
  We will call this solution  ``quasi-stationary" where by  ``quasi'' we mean that it is an approximate solution, valid as long as vector $\vec{\Omega}_c$ can be treated as a constant.
  The general solution of homogeneous equations is a sum of modes proportional to $\exp(pt)$.
  It is not difficult to see that they are the same modes as were obtained in Section \ref{sec:three_comps}  (except the long-period precession mode). 
  Indeed, making the following formal change of variables:
  \begin{align}
    &\nu_{gc}^{(\parallel)} = \mu_{g}^\parallel - \mu_{c}^\parallel, \ \ \
    \nu_{rc}^{(\parallel)} = \mu_{r}^\parallel - \mu_{c}^\parallel, \\
    &\nu_{gc}^{(\perp)} = \mu_{g}^\perp - \mu_{c}^\perp,  \ \ \
    \nu_{rc}^{(\perp)} = \mu_{r}^\perp - \mu_{c}^\perp, \\
    & \frac{\omega_\mathrm{sf}^{(\perp)}}{\omega_\mathrm{sf}^{(\parallel)}} = \frac{\omega_\mathrm{sf}^{\perp}}{\omega_\mathrm{sf}^{\parallel}} - \frac{\mu_{c}^\perp}{\Omega_0},
  \end{align}
  and renaming $\Omega_0$ to $\Omega_c$,
  we can obtain system of equation  \eqref{eq:dnu_gc_par_quasistat} -- \eqref{eq:dom_sf_perp_quasistat} with zeroth $\vec{S}$ from system of equations  \eqref{eq:dmu_g_par_three_comps} -- \eqref{eq:dom_sf_perp_three_comps} with zeroth $\epsilon_c$.}
  Since all the real parts of modes \eqref{eq:par_mode_+_three_comps}, \eqref{eq:par_mode_-_three_comps} and  \eqref{eq:rot_mode_1_three_comps} -- \eqref{eq:prec_mode_g_three_comps} are negative, small perturbations decay to the quasi-stationary solution.
 
  The quasi-stationary values of $\nu_{gc}^{(\parallel)}$, $\nu_{rc}^{(\parallel)}$, ${\nu}_{gc}^{(\perp)}$, ${\nu}_{rc}^{(\perp)}$ and ${\omega}_\mathrm{sf}^{(\perp)}$ can be obtained from equations \eqref{eq:dnu_gc_par_quasistat} -- \eqref{eq:dom_sf_perp_quasistat} with simple algebra. 
  The expressions, however, are quite cumbersome.
  We do not give exact expressions, restricting ourselves to estimations. 
  Let us first look at the ``parallel" part of the system of equations. 
  Since the time derivatives are negligibly small, the non-zeroth value of $S^{(\parallel)}$ can be counterbalanced only by interaction torques $R_{ij}^{(\parallel)}$,
  i.e. $R_{ij}^{(\parallel)}\sim S^{(||)}$ and $\nu_{ij}^{(\parallel)} \sim S^{(||)} /\mathrm{min}(\alpha_{jk})$.
  The precession term does not contribute to $S^{(\parallel)}$. 
  Therefore, $S^{(\parallel)} \sim K_3/I_c \sim(I/I_c) (\Omega_c/\tau_c)$ where $\tau_c$ is the characteristic age of the pulsar.
  For the ``perpendicular" part we have ${\nu}_{ij}^{(\perp)}, \ {\omega}_\mathrm{sf}^{(\perp)}\sim {S}^{(\perp)}/\Omega_c$.
  The leading term in ${S}^{(\perp)}$ is the precession term.
  Hence, ${\nu}_{ij}^{(\perp)}, \ {\omega}_\mathrm{sf}^{(\perp)} \sim \epsilon_c \Omega_c$.
  Thus, we see that the quasi-stationary values of these quantities are small as it was assumed.

  { Let us turn to  the long-period precession time-scale.
  If the small perturbations are not excited during stellar evolution, they  should all be damped on this time-scale.}   
  However, we obviously can no longer ignore the evolution of vector $\vec{\Omega}_c$.
  It produces corrections to the quasi-stationary expressions arising to compensate the time derivatives in equations \eqref{eq:dnu_gc_par_quasistat} -- \eqref{eq:dom_sf_perp_quasistat}.
  The time derivatives, in turn, arise due to changing the components of vector $\vec{\Omega}_c$ contained in the quasi-stationary expressions. 
  Hence, $\dot{\nu}_{ij} \sim \nu_{ij} (\dot{\Omega}_c/\Omega_c)$.
  According to equations \eqref{eq:dOm_c_par_quasistat} and  \eqref{eq:dOm_c_perp_quasistat} we have $\dot{\Omega}_c \sim \alpha_{ij} \nu_{ij}$ and ${\dot{e}}_\Omega \sim \nu_{ij}$.
  Hence, $\dot{\nu}_{ij} \sim \nu_{ij}^2$, i.e. the corrections are quadratically small. 
  Thus,  we can continue to use the quasi-stationary expressions for small departures at the long-period time-scales if the three following conditions are satisfied.
  First, the departures from the solid-body rotation should be small. 
  Secondly, the internal relaxation time-scale should be much smaller than the slow precession period $T_p$ (ensured by conditions \eqref{eq:qusistat_ineq_val}).
  {Thirdly, the time-dependent perturbations are not excited during the star's life-time.
  The last condition is obviously violated if the neutron star is glitching.
  This case is the subject of the next section.}
  
  The quasi-stationary approximation allows us to exclude the  rotation of the internal components from consideration.
  Let us first look at the ``parallel" part.
  The quasi-stationary expressions for  $\nu_{gc}^{(\parallel)}$ and $\nu_{rc}^{(\parallel)}$, being expressed from equations \eqref{eq:dnu_gc_par_quasistat} and \eqref{eq:dnu_rc_par_quasistat}, can be substituted into the right-hand side of equation \eqref{eq:dOm_c_par_quasistat}.
  However, a more illustrative way to obtain the same result is the following. 
   Taking a look at equations \eqref{eq:dnu_gc_par_quasistat} and \eqref{eq:dnu_rc_par_quasistat}, we notice that the three last terms on their right-hand sides (cf. equation \eqref{eq:dOm_c_perp_quasistat}) are exactly equal to $\dot{\Omega}_c$. 
  After transposing $\dot{\Omega}_c$ to the left-hand sides of the equations one can compose the combination $I_c$\eqref{eq:dOm_c_par_quasistat}$+ (1+\epsilon_g)I_g$\eqref{eq:dnu_gc_par_quasistat}$+ I_r$\eqref{eq:dnu_rc_par_quasistat}.
  Neglecting quadratically small time derivatives $\dot{\nu}_{ij}^{(\parallel)}$ and taking into account relations \eqref{eq:R_ij_relation}, we finally obtain
  \begin{equation}
    \label{eq:braking}
    \tilde{I}\dot{\Omega}_c = K_3^{(\parallel)},
  \end{equation}
  where we have denoted $\tilde{I} = I_c+(1+\epsilon_g)I_g + I_r$.
  According to this equation the neutron star is braked by the external torque as if it is a rigid body with the moment of inertia equal to $\tilde{I}$.
  It  does not contain the moment of inertia of the pinned superfluid because the superfluid does not slow down as long as it remains pinned.
  The rigid-body braking is a general feature of the quasi-stationary approximation \citep[section 2.3]{BGT2014}.
    
  Let us turn to the ``perpendicular" part.  
  According to equation  \eqref{eq:dOm_c_perp_quasistat}, the  right-hand side of equation \eqref{eq:dom_sf_perp_quasistat} is equal to $-\dot{{e}}_\Omega$.
   Therefore, from equation \eqref{eq:dom_sf_perp_quasistat} in the quasi-stationary approximation ($\dot{{\omega}}_\mathrm{sf}^{(\perp)}\approx 0$)  we have
  \begin{equation}
    \label{eq:nu_gc_perp_quasistat}
    {\nu}_{gc}^{(\perp)} = i \dot{{e}}_\Omega.
  \end{equation}
  Substituting equation \eqref{eq:nu_gc_perp_quasistat} into equations \eqref{eq:dnu_gc_perp_quasistat} and \eqref{eq:dnu_rc_perp_quasistat} (with zeroth time derivatives), we obtain the system of linear equations for  $\nu_{rc}^{(\perp)}$ and $\omega_\mathrm{sf}^{(\perp)}$, allowing us to express these quantities as functions of $\dot{{e}}_\Omega$. The solution is the following:
  \begin{align}
    \label{eq:nu_rc_perp_quasistat}
    &\nu_{rc}^{(\perp)} = i \dot{e}_\Omega - \frac{ \xi_{cr}}{ \Omega_c - i  \xi_{cr} - i  \xi_{gr} }\dot{e}_\Omega, \\
    \label{eq:omega_sf_perp_quasistat}
    &\omega_\mathrm{sf}^{(\perp)} = i \dot{e}_\Omega + \frac{\dot{e}_
    \Omega}{\tilde{\epsilon}_g}\left( \frac{\xi_{cg}}{\Omega_c} + i \frac{\xi_{rg}}{\Omega_c} \frac{   \xi_{cr}}{ \Omega_c - i  \xi_{cr} - i  \xi_{gr}} \right).
  \end{align}  
  We see that all three quantities \eqref{eq:nu_gc_perp_quasistat} -- \eqref{eq:omega_sf_perp_quasistat} contain the same term $i \dot{{e}}_\Omega$. 
  The last two contain some corrections to it that, however, are small in the weak-interaction limit.
  Substituting equations \eqref{eq:nu_gc_perp_quasistat} and \eqref{eq:nu_rc_perp_quasistat} into equation \eqref{eq:dOm_c_perp_quasistat} and making some rearrangement, we finally obtain the equation for $\dot{e}_\Omega$, which after returning to the vector form can be represented as
    \begin{align}
      \nonumber
      I_c \Omega_c \dot{\vec{e}}_{\Omega} =- (1+\Gamma) \frac{\vec{e}_\Omega\times\left[\vec{e}_\Omega\times\vec{K}_3\right] + I_c\vec{\Omega}_c\times\tensor{\epsilon}\vec{\Omega}_c}{(1+\Gamma)^2+B^2}  \\
     \label{eq:precess} 
      +B \frac{\vec{e}_\Omega\times(\vec{K}_3-I_c\vec{\Omega}_c\times\tensor{\epsilon}\vec{\Omega}_c)}{(1+\Gamma)^2+B^2},
    \end{align}  
    where coefficients $B$ and $\Gamma$ are determined by equality
    \begin{equation}
      \label{eq:precess_coeffs}
       B + i \Gamma = \xi_{gc} + \xi_{rc} + \frac{\xi_{rc} \xi_{cr} }{\Omega_c - i
       \xi_{cr} - i  \xi_{gr} }.
     \end{equation}
     Here we have used notation 
     \begin{equation}
      \label{eq:epsilon_axisym}
       \tensor{\epsilon}\vec{\Omega}_c = \epsilon_c \vec{e}_c (\vec{e}_c\cdot\vec{\Omega}_c).
      \end{equation}
     However,  all steps of the derivation of equation \eqref{eq:precess} remain valid if we assume a more general form for the c-component angular momentum
     \begin{equation}
       \vec{M}_{c} = I_{c} \vec{\Omega}_c + I_{c} \tensor{\epsilon} \vec{\Omega}_c
     \end{equation}
     instead of form \eqref{eq:Mc_three_comps}, where $\tensor{\epsilon}$ is an arbitrary symmetric tensor describing small deviations of the c-component from spherical symmetry.
     
  Since the weak-interaction limit is assumed ($|\xi_{ij}|/\Omega_c\ll1$), the coefficients in equation \eqref{eq:precess}  approximately are equal to 
  \begin{equation}
    \label{eq:precess_coeffs_weak_limit}
    B \approx \frac{\beta_{gc} + \beta_{rc}}{\Omega_c}, \ \ \ \ \Gamma \approx \frac{\gamma_{gc} + \gamma_{rc}}{\Omega_c}.
  \end{equation}     
  Replacing expression \eqref{eq:precess_coeffs} by expressions \eqref{eq:precess_coeffs_weak_limit}, we ignore the difference between the rotation of the two internal components (the second term in equation \eqref{eq:nu_rc_perp_quasistat} is neglected).
  
  If the internal components do not interact with the c-component at all, both coefficients $B$ and $\Gamma$ are equal to zero. In this case, equation \eqref{eq:precess} takes the form 
  \begin{equation}
    I_c \Omega_c \dot{\vec{e}}_{\Omega} =-  \vec{e}_\Omega\times\left[\vec{e}_\Omega\times\vec{K}_3\right]  - I_c\vec{\Omega}_c\times\tensor{\epsilon}\vec{\Omega}_c.
    \end{equation}
  With $\vec{K}_3 = 0$ and equality \eqref{eq:epsilon_axisym} it reproduces the rigid-body free precession equation \eqref{eq:dOm_rigid_body}. 
  The influence of $\vec{K}_3$ at the precession evolution is beyond the scope of the present paper. This question has already been well studied \citep{Goldreich1970,Melatos2000,ArzamasskiyEtAl2015,GBT2015}. 
  The electromagnetic torque can force the precession amplitude to both increase  or decrease at the time-scales of pulsar characteristic age $\tau_c$.
  It is difficult to formulate any more specific statement because the result depends strongly on the mutual orientation of the stellar magnetic axis and  its principal axes as well as on the exact form of the external electromagnetic torque $\vec{K}_3$. 
  
  Next, let us switch off the external torque but take into account the interaction with the internal components. 
  If $\vec{K}_3 = 0$, according to equation \eqref{eq:braking} we have $\dot{\vec{\Omega}}_c = \Omega_c\dot{\vec{e}}_{\Omega}$.
  Since the interaction is assumed to be weak, we keep only the linear in $B$ and $\Gamma$ terms.
  The equation in the axisymmetric case would be the following:
  \begin{equation}
    \label{eq:precess_wo_K3}
    \dot{\vec{\Omega}}_c = \epsilon_c(\vec{e}_c\cdot\vec{\Omega}_c)\left\{(1-\Gamma) [\vec{e}_c\times\vec{\Omega}_c]+ B \vec{e}_\Omega\times [\vec{e}_c\times\vec{\Omega}_c]\right\}.
  \end{equation}
  The interaction with the internal components modifies the rigid-body precession equation \eqref{eq:dOm_rigid_body} in two ways.
  Coefficient $\Gamma$ produced by non-dissipative interaction renormalizes the precession period {(cf. with expression \eqref{eq:Tp_rigid_body}):
  \begin{equation}
    \label{eq:Tp_withGamma}
    T_p \approx \frac{P}{(1-\Gamma) \epsilon_c  \cos\theta}.
  \end{equation}
  }
  {It is easy to see that in the small-angle approximation}  this coincides with value \eqref{eq:Tp_three_comps} obtained in the framework of the linear analysis.
  Considering the influence of coefficient $B$, let us multiply equation \eqref{eq:precess_wo_K3} by  vector $\vec{e}_c$:
  \begin{equation}
    \label{eq:align}
    (\dot{\vec{\Omega}}_c\cdot\vec{e}_c) = \Omega_c d_t \cos\theta =   \epsilon_c B \Omega_c^2\cos\theta\sin^2\theta  \end{equation}
  Here, $(\vec{e}_c\cdot\vec{\Omega}_c)>0$ by choosing the direction of $\vec{e}_c$; coefficient $B$ calculated with formula \eqref{eq:precess_coeffs_weak_limit} is positive because interaction \eqref{eq:R_quasistat} should reduce the difference between the rotation the components.
  Therefore, angular velocity $\vec{\Omega}_c$ tends to align with the symmetry axis $\vec{e}_c$.
  The characteristic time-scale of the alignment is 
  {
  \begin{equation}
    \label{eq:tau_align_quasistat}
    \tau_\mathrm{align}  \sim \frac{T_p}{2\pi B}.
  \end{equation}
  } 
  This is the same time-scale that we obtained for $p_3$ mode in Section \ref{sec:three_comps}.
  
  The quasi-stationary evolution formalism confirms the results obtained in the framework of the linear mode analysis for the more general case of an arbitrary angle between the angular velocity and the c-component symmetry axis.
  However, the main advantage of this formalism is that it allows us to obtain equations \eqref{eq:braking} and \eqref{eq:precess}, which can be used for studying the stellar rotation evolution at the time-scales comparable to the pulsar life time.

\section{Glitch-like event}
  \label{sec:glitch} 
  We have formulated a model of a rotating neutron star with pinned superfluid that allows the neutron star to precess with long period.
  The model should also be examined for the ability to demonstrate glitch-like behaviour.

  {
   We assume that the glitch relaxation (the longest glitch stage) is governed by internal relaxation processes.
   As it was discussed in the  previous section, at internal relaxation time-scale $\tau_r$,  vectors $\vec{\Omega}_c$ and $\vec{S}$ can be treated as constants in equations \eqref{eq:dnu_gc_par_quasistat} -- \eqref{eq:dom_sf_perp_quasistat}.
   In this case, the full solution can be represented as a sum of the quasi-stationary  solution (considered in the previous section) and a time-dependent solution to the homogeneous system of equations (with zeroth $S^{(\parallel)}$ and $S^{(\perp)}$). The last, in turn, is a sum of the linear modes studied in Section \ref{sec:three_comps} (except the $p_3$ mode, which has already been taken into account in the quasi-stationary solution; see Section \ref{sec:qstat_evolution}).
   Generally speaking, the glitch can excite both ``parallel" and ``perpendicular" modes. 
   However, in the present paper, we will consider only the ``parallel" perturbations, which allow us to reproduce the main glitch manifestation --  the pulsar rotation  frequency jump. 
  Thus, the system of equations is reduced to 
  \begin{align}
    \label{eq:dOm_c_par_glitch} 
    &\dot{\Omega}_c   = \alpha_{gc}\nu_{gc}^{(\parallel)}+\alpha_{rc}\nu_{rc}^{(\parallel)},\\
    \label{eq:dnu_gc_par_glitch}
    & {\dot{\nu}_{gc}^{(\parallel)}}  = 
      - \left( \alpha_{gc} +\frac{\alpha_{cg}}{1+\epsilon_g} + \frac{\alpha_{rg}}{1+\epsilon_g}\right) \nu_{gc}^{(\parallel)}  \\ 
      & \ \ \ \ \ \ \ \ -\left(\alpha_{rc} -  \frac{\alpha_{rg}}{1+\epsilon_g}\right)\nu_{rc}^{(\parallel)} \nonumber
    \\
    \label{eq:dnu_rc_par_glitch}
    & {\dot{\nu}_{rc}^{(\parallel)}}  =- \left( \alpha_{rc} +\alpha_{cr} + \alpha_{gr}\right) \nu_{rc}^{(\parallel)} - \left(\alpha_{gc} -  \alpha_{gr}\right)\nu_{gc}^{(\parallel)}.
  \end{align}         
   
   The exact glitch-triggering mechanism is not known at present \citep{HaskellMelatos2015}.  
   The mechanism is likely based on some non-linear process that cannot be described by simple linear equations.
   We will model it in the following way.
   It is assumed that at $t=0$ the  superfluid pinned in the g-component instantly releases a small part of the stored angular momentum $\Delta \vec{L}_\mathrm{sf}$. 
   Since we are interested in studying the ``parallel" modes,  released angular momentum $\Delta \vec{L}_\mathrm{sf}$ is supposed to be parallel to $\vec{\Omega_c}$.
   The whole released angular momentum is assumed to be instantly injected into the ``normal" fraction of the g-component such that the total g-component angular momentum is conserved.
  Thus, at $t=0$  the value of $\omega_\mathrm{sf}^{(\parallel)}$ decreases by $\Delta L_g/I_g$ while angular velocity $\Omega_g$ increases by $\Delta L_g/I_g$.  
  The value of $\omega_\mathrm{sf}^{(\parallel)}$ is not contained in  equations \eqref{eq:dOm_c_par_glitch} -- \eqref{eq:dnu_rc_par_glitch}. 
  Thus, right after the triggering we have:   $\Omega_c = \Omega_0$, $\nu_{rc}^{(\parallel)} = 0$, $\nu_{gc}^{(\parallel)} = \Delta L_g/I_g$.   
  We can consider these values as initial conditions and study how the system evolves.
  }
  
\begin{figure}
          \center
          \includegraphics[width=0.4\linewidth,trim= 0.4cm 1.05cm 0.6cm 0.cm, clip = true]{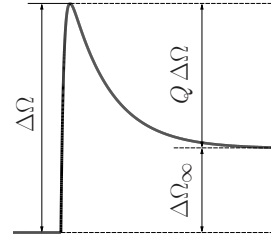}
          \caption{\label{fig:glitch}A sketch of the glitch-like behavior of $\Omega_c$.}        
  \end{figure}

  {
  To find the c-component rotation response to the glitch in the g-component
  we can substitute the solution to equations \eqref{eq:dnu_gc_par_glitch} -- \eqref{eq:dnu_rc_par_glitch} 
  into equation \eqref{eq:dOm_c_par_glitch} and then integrate it.
  The result can be represented in the following form:}
  \begin{equation}
      \label{eq:Omega_glitch}
      \Omega_c(t) = \Omega_0 + \Delta\Omega
      \left( 1 - e^{p_{+}t} - Q (1-e^{p_{-}t}) \right),
  \end{equation}
  where
  $\Delta\Omega = {\Delta \Omega_{\infty}}/{(1-Q)}$,
  $\Delta\Omega_{\infty} = { \Delta L_{g}}/{\tilde{I}}$,
  \begin{equation}
    \label{eq:Q}
    Q = \frac{\tilde{I} \alpha_{gc} + (1+\epsilon_g)I_{g} p_{+} }{ \tilde{I} \alpha_{gc} +(1+\epsilon_g) I_{g} p_{-} }
  \end{equation}
   and
  coefficients $p_{+}$ and $p_{-}$ are the two ``parallel" modes.
  If one wants to relate the solution  to observed pulsar glitches, function \eqref{eq:Omega_glitch} should have the form shown in Fig. \ref{fig:glitch}.
  Values $1/p_{+}$ and $1/p_{-}$ should be interpreted as glitch rise
  and glitch relaxation 
  time-scales respectively.
  Hence, it is required  that $p_+ \gg p_-$.
  This condition is satisfied if one of the pair interactions is much stronger than the other two (see Section \ref{sec:three_comps}).
  {Let us first assume that assume that the strongest one is the cg-interaction, i.e. inequality \eqref{eq:sigma_cond} is satisfied.}
  In this case, coefficients $p_+$ and $p_-$ can be calculated with expressions \eqref{eq:par_mode_+_three_comps} and \eqref{eq:par_mode_-_three_comps}.
  If we want to reproduce the angular velocity behaviour sketched in Fig. \ref{fig:glitch}, we should also ensure that \citep{LyneShemarSmith2000}
  \begin{equation}
    \label{eq:Q_cond}
    0<Q<1.
  \end{equation}  
  Substitutting expressions \eqref{eq:par_mode_+_three_comps} and \eqref{eq:par_mode_-_three_comps} into expression \eqref{eq:Q} we obtain
  \begin{equation}
    \label{eq:Q_est}
    Q \approx \frac{I_r}{\tilde{I}}.
  \end{equation}
  If we choose another interaction to be the strongest one assuming, for instance, that $\sigma_{cr} \gg \sigma_{cg}, \sigma_{gr}$,  the estimations for        $p_+$ and  $p_-$ can be obtained from expressions \eqref{eq:par_mode_+_three_comps} and \eqref{eq:par_mode_-_three_comps} by the corresponding  interchange of indices.
  However, condition \eqref{eq:Q_cond}, {in this case, requires fine tuning of the interaction parameters, which is unlikely to be able to be maintained for long time since the parameters evolve with the neutron star's internal temperature. }
 Thus, the case of inequality \eqref{eq:sigma_cond} is most plausible. 
 
 From the physical point of view the picture is as follows. 
 The glitch suddenly increases the  g-component { ``normal" fraction} angular velocity making angular velocity lag $\nu_{gc}^{(\parallel)}$ exceed its quasi-stationary value.
 Right after that the c-component spins up due to the cg-interaction at the  (observationally unresolved) $p_+$-time-scale.  
 The spinning up is followed by spinning down due to the reaction of the r-component.
 Thus, in the framework of the proposed model, the glitch  relaxation is provided by the presence of the r-component. 
 Indeed, putting $I_r=0$ in expression \eqref{eq:Q_est}, we obtain $Q=0$ and, hence, $\Delta\Omega_{\infty} = \Delta\Omega$.
 
 {Observations could give us some constraints for $p_+$ and $p_-$ and, hence, for the interaction coefficient parameters.
 From expression \eqref{eq:par_mode_+_three_comps} it is follows that
 \begin{equation}
    \label{eq:sigma_cg_glitch}
    \sigma_{cg} \sim \ \frac{P}{2\pi \tau_\mathrm{glitch}},
  \end{equation}
  where $\tau_\mathrm{glitch}$ is the  characteristic glitch rise time. 
 Supposing that $\beta_{ij}\sim\alpha_{ij}$ and taking into account condition \eqref{eq:sigma_cond}, we can relate $\sigma_{cg}$ to  $B$:
 \begin{equation} 
   \label{eq:sigma_cg_B}
    \sigma_{cg} \sim \frac{I_c +(1+\epsilon_g)I_g}{(1+\epsilon_g)I_g} B.
 \end{equation}
 
 The relaxation time-scale $\tau_\mathrm{g.relax.}$ allows us to estimate the other interaction coefficients. From expression \eqref{eq:par_mode_-_three_comps}, we have 
 \begin{equation}
   \label{eq:alpha_relax-time}
   \alpha_{cr} + \alpha_{gr} \sim \frac{I_c +(1+\epsilon_g)I_g}{\tilde{I}} \tau_{g.relax.}
 \end{equation}
 }
 

\section{Discussion} 
\label{sec:discussion}
  {We have formulated a simple model of a rotating neutron star 
  supposing the existence of three abstract dynamically distinguished components.
  }
  Let us now speculate on the possible physical counterparts of these components.
  
  The outer c-component can be represented by the neutron star's crust and  part of its core  strongly coupled with the crust. 
  Therefore,  the moment of inertia of the c-component can be estimated as $I_c\sim(10^{-2}-10^{-1})I$. 
  
  The role of the g-component can be performed by tangles of closed flux tubes that could be formed  from chaotic small-scale magnetic field after protons became superconductive.
     Alternatively it could be a torus composed of closed flux tubes \citep{GugercinougluAlpar2014}. If, for instance, the characteristic cross-section $S_\mathrm{tor}$ of the region occupied by the toroidal field is of the order of 1 km$^{2}$, then $I_g\sim \rho_p S_\mathrm{tor} r_\mathrm{ns}^3 \sim 10^{-3}I$, where $\rho_p$ is the proton mass density.
  Some of the superfluid neutron vortices located in the core can be pinned to the closed flux tubes. 
  On the one hand, this interaction prevents the tangles or torus collapsing. On the other hand, when the critical rotational lag is reached, the vortices unpin, triggering the glitch.
  These pinned vortices carry angular momentum $\vec{L}_\mathrm{sf}$. 
  {If we assume that  1 percent of the total stellar moment of inertia is contained in the pinned superfluid,  then ${L}_\mathrm{sf}/I_g\Omega_0 \approx 10^{-2} (I/I_g)$.
  Assuming that a thin ring is a good approximation for the g-component mass distribution, the real oblateness parameter can be estimated as $\epsilon_g \approx 2$.
  Hence, the second term in expression \eqref{eq:epsilon_g_three_comps} is dominant and $\tilde{\epsilon}_g \sim 10^{-2} (I/I_g) \sim 10$.
  }
  
  Since the flux tubes are closed in the core, the g-component, being magnetically decoupled, can rotate with an angular velocity different from $\vec{\Omega}_c$ \citep{GlampedakisLasky2015}.   
  As it has been argued, the cg-interaction should be quite strong to ensure the rapid c-component spin up but not too strong to allow the g-component  to {rotate} relative to the crust.   
  We guess that this interaction can be implemented by the viscosity of crust-core interface region with the possible inclusion of the Ekman pumping mechanism.
  { Unfortunately, there are not many observational data  for glitch rise times at present.
  The Crab pulsar is the only pulsar for which partially resolved spin-up data have been obtained \citep{ShawEtAl2018}.
  The delayed frequency increase lasted for 0.5--1.7 days. However, this phenomenon was observed only in three large Crab glitches. 
  Therefore, it could be more correct to apply the upper limit for the fast unresolved part of the glitch spin-up for estimating the cg-interaction coefficients.
  For the  2017 glitch \citet{ShawEtAl2018} placed the upper limit $\tau_\mathrm{glitch}<6$ h.  Hence, for the Crab pulsar with $P=3.3\times10^{-2}$ s \citep{ATNF_paper}\footnote{http://www.atnf.csiro.au/research/pulsar/psrcat/}, 
  using estimation \eqref{eq:sigma_cg_glitch},
  we obtain  $\sigma_{cg}  >  2.4\times10^{-7}$.
 For the Vela pulsar much stiffer restrictions were obtained. For 2004 glitch, \citet{DodsonLewisMcCulloch2007} obtained that $\tau_\mathrm{glitch}<30$s 
  and, hence, $\sigma_{cg} > 4.7\times 10^{-4}$, where $P=8.9\times10^{-2}$ s for the Vela pulsar\footnotemark[2].
 The value of $\sigma_{cg}$ can vary substantially from one pulsar to another due to different internal temperatures and magnetic flux tube organization. 
 Moreover, we do not state that the presented model describes all glitching pulsars. In some of them the g- and the c-component could be tightly coupled. This part, however, cannot precess with long periods, as it was shown earlier. 
  }


  {From the observed glitch relaxation time-scales \citep{LyneShemarSmith2000}, using expression \eqref{eq:alpha_relax-time}, we obtain the following estimation for the corresponding interaction parameters:
 \begin{equation}
    \alpha_{cr} + \alpha_{gr} \sim \frac{I_c +I_g}{\tilde{I}} (1 - 10^2) \mbox{ days}.
 \end{equation}
 }
  The weakly interacting r-component is likely composed of superfluid matter. 
  We suppose this component consists of some of the neutrons superfluid that is not pinned
  {and possibly some of the normal matter coexisting with it and  weakly coupled with c-component.} 
  Hence, in order of magnitude, we can estimate that $I_r\sim I$.
   {Strictly speaking, it would be more consistent to treat the superfluid and normal fractions of the r-component as  two separate components because of their weak interaction. It would complicate the calculations and would probably lead to a more complex glitch relaxation but qualitatively the model remains the same.}
   
   It is generally accepted that the neutrons in the inner crust should be superfluid and, hence, the superfluid vortices can pin to the crust lattice.
   According to theoretical calculations, about 1 percent of the total moment of inertia is contained in the crust superfluid. If all that superfluid is pinned, the crust's effective oblateness parameter would be of the order of $10^{-2}$ and, hence, long-period precession becomes impossible. 
   However, as  shown by \citet{LinkCutler2002}, the Magnus force, acting on the crust vortices could be enough to unpin them.
   Thus, we assume that the crust superfluid is unpinned in precessing stars. Strictly speaking, in this case, it should be included in the model as an additional component. Again, it can modify coefficients $B$ and $\Gamma$ and complicate the glitch relaxation. The exact effects of mentioned additional components are the subject of future study. 
   
  { 
  \citet{AshtonJonesPrix2017} found that the modulations in spin-down rate and beam shape of pulsar B1828-11 become faster.  If the modulations are induced by precession,  the rate of precession period decrease is $\dot{T}_p \approx - 10^{-2} \mbox{s s}^{-1}$. According to expression \eqref{eq:Tp_rigid_body} this may indicate the gradual increase of the star's oblateness, which would be quite counter-intuitive.
  Alternatively, the variation of precession period could be caused by a decrease in angle $\theta$. In the case of a rigidly rotating star, the angle evolution can be caused only by electromagnetic torque $\vec{K}$; this process would occur on the same time-scale as the pulsar spin-down and, hence, it cannot be the source of such  fast variation.
  However, the pulsar could align due to internal dissipation. Taking the time derivative of precession period \eqref{eq:Tp_withGamma} and using equation \eqref{eq:align}, we can obtain the estimation for the corresponding coefficient:
  \begin{equation}
    \label{e:B_dTp}
    B \approx - \frac{\dot{T}_p}{2\pi} \frac{\cos\theta}{\sin^2\theta}.
  \end{equation}
  Since the spin-down rate is not affected by internal dissipation (see equation \eqref{eq:braking}) in the quasi-stationary approximation, the rotation frequency can be treated as a constant at the considered time-scale.
   Due to the symmetry of the problem there are two configurations  with $\theta = 5^\circ$ and $\theta = 89^\circ$, which fit the observational data equally well  \citep{ArzamasskiyEtAl2015}. 
   For the small angle configuration with estimations \eqref{e:B_dTp} and \eqref{eq:sigma_cg_B} we formally obtain $B \approx 0.2$ and $\sigma_{cg} \sim 1 - 10 $. This case is not self-consistent because the weak-interaction condition is not satisfied. 
   In the case of the large angle configuration we have $B \approx 3\times10^{-5}$ and $\sigma_{cg} \sim 10^{-4} - 10^{-3}$. This, according to estimation \eqref{eq:sigma_cg_glitch}, leads for the pulsar with $P=0.4$ s to $\tau_\mathrm{glitch} \sim 10^2 - 10^3$ s. 
   These values seem to be quite plausible.
   The configuration with large angle $\theta$ assumes a small angle between the symmetry axis and the magnetic dipole axis. 
   Therefore, speculatively we can guess that, some time in the past, a starquake  occurred, which led to the rearrangement of the crust almost along the magnetic axis. This occasion started the precession.
   The phenomenon of decreasing $T_p$, however, could have an alternative explanation. For example, the crust superfluid, which is unpinned due to precession, can gradually re-pin, which caused the increase of the crust's effective oblateness.  
  } 
  
  {
    In the present paper we proposed a way in which the vortex-pinning-based pulsar glitches can be reconciled with the long-period precession.  Basically it is assumed that the superfluid is pinned in the region located in the  stellar interior, which  has sufficient freedom to rotate relative to the crust. This allows the pinned superfluid's angular momentum to stay aligned almost along the stellar rotation axis and, hence, to not affect the precession period.
    At the same time, this region should be sufficiently effectively coupled with the crust to ensure the sufficiently rapid spin-up of the crust after a glitch there has happened. 
   Since the precession and the glitches are governed by different regions of the star, it is natural that the glitch has almost no effect on the precession characteristics \citep{JonesAshtonPrix2017}. 
   The exact glitch-triggering mechanism, however, is beyond the scope of the model in its present form. It is  rather brought into it by hand.
   Therefore, the model by itself does not allow us to predict the sizes and the waiting times of the glitches but these questions could be addressed in the existing core-located glitch models. The main new point that we assume is the possible weakness of cg-interaction. This assumption does not affect the ``normal" fraction spin-down relative to the pinned superfluid,  which is usually supposed to be the key process for glitch triggering.
    }


  Unfortunately, our  model in its present form does not allow us to obtain the whole range of observed recovery fraction values. According to estimation \eqref{eq:Q_est} it is of the order of unity. It is not so bad for young pulsars but 
  the older ones demonstrate a wide range of recovery parameters 
  \citep{YuEtAl2013}. However, we believe that this discrepancy arises due to oversimplification of the model.
  {In particular, the injection of the angular momentum from the pinned superfluid into the ``normal" fraction requires more self-consistent consideration.
   Thus,  further research is required.}

 \section*{Acknowledgments}
The authors are grateful to A I Tsygan, K Y Kraav, M V Vorontsov, and A N Yurkin for fruitful discussions. The authors are also grateful to Dr. Ashton for many useful comments.

\bibliographystyle{mnras}
\bibliography{mn-jour,paper}





\bsp	
\label{lastpage}
\end{document}